\documentclass{aa}  

\usepackage{hyperref}
\usepackage{graphicx, subcaption, color, float}
\usepackage{xcolor}
\definecolor{darkblue}{rgb}{0,0,0.5} 
\usepackage{transparent}
\usepackage[adobe-utopia]{mathdesign}
\usepackage{amsmath, amssymb, bm, gensymb}
\usepackage{mathtools}

\usepackage[boxed]{algorithm2e}

\SetCommentSty{mycommfont}

\usepackage{wrapfig}

\allowdisplaybreaks[3]

\newcommand{\bs}[1]{\ensuremath{\boldsymbol{#1}}}
\renewcommand{\vec}[1]{\ensuremath{\mathbf{#1}}}

\newcommand{\g}{\ensuremath{\vec{g}}}

\newcommand{\x}{\ensuremath{\vec{x}}}

\newcommand{\Z}{\ensuremath{\vec{Z}}}

\newcommand{\J}{\ensuremath{\vec{J}}}

\newcommand{\E}{\ensuremath{\vec{E}}}
\newcommand{\ddtec}{\ensuremath{\Delta_0^2\tau}}
\renewcommand{\k}{\ensuremath{\vec{k}}}

\newcommand{\I}{\ensuremath{\vec{I}}}
\newcommand{\B}{\ensuremath{\vec{B}}}
\newcommand{\y}{\ensuremath{\vec{y}}}

\renewcommand{\iota}{\textit{i}}

\DeclareMathOperator*{\argmax}{argmax}

\usepackage{mathtools}

\DeclarePairedDelimiter\floor{\lfloor}{\rfloor}

\begin{document}

\title{Probabilistic direction-dependent ionospheric calibration for LOFAR-HBA}

\author{J.~G.~Albert\inst{1} \and R.~J.~van~Weeren\inst{1}  \and H.~T.~Intema\inst{1, 2} \and H.~J.~A.~R\"ottgering\inst{1}}

\institute{1. Leiden Observatory, Leiden University, P.O. Box 9513, 2300 RA Leiden, the Netherlands\\
2. International Centre for Radio Astronomy Research -- Curtin University, GPO Box U1987, Perth WA 6845, Australia\\
E-mail: \href{mailto:albert@strw.leidenuniv.nl}{albert@strw.leidenuniv.nl} }

\date{Received December 30, 2019 / Accepted January 31, 2020}

\abstract{
Direction dependent calibration and imaging is a vital part of producing deep, high fidelity, high-dynamic range radio images with a wide-field low-frequency array like LOFAR.
Currently, state-of-the-art facet-based direction dependent calibration algorithms rely on the assumption that the isoplanatic-patch size is much larger than the separation between bright in-field calibrators.
This assumption is often violated due to the dynamic nature of the ionosphere, and as a result direction dependent errors affect image quality between calibrators.
In this paper we propose a probabilistic physics-informed model for inferring ionospheric phase screens, providing a calibration for all sources in the field of view. 
We apply our method to a randomly selected observation from the LOFAR Two-Metre Sky Survey archive, and show that almost all direction dependent effects between bright calibrators are corrected and that the root-mean-squared residuals around bright sources is reduced by 32\% on average.
}


   \maketitle

%________________________________________________________________

\section{Introduction}

In radio astronomy many of the biggest questions are only answerable by observing the faintest emission over a large fraction of the sky surface area.
Such big puzzles include searching for the epoch of reionisation \citep[e.g.][]{2017ApJ...838...65P}, potentially detecting the missing baryons in the synchrotron cosmic web \citep[e.g.][]{Vernstrom2017}, understanding the dynamics of galaxy cluster mergers \citep[e.g.][]{2019SSRv..215...16V}, and probing matter distribution with weak gravitational lensing at radio frequencies \citep[e.g.][]{2016MNRAS.463.3674H}.
In recent years, instruments like the Low-Frequency Array \citep[LOFAR;][]{2013A&A...556A...2V}, the Murchison Widefield Array (MWA), and the future Square Kilometre Array (SKA) have been designed with the hopes of discovering these key pieces in our understanding of the universe.
In particular, the prime objective of LOFAR, a mid-latitude array (N$52^\circ$), is to be a sensitive, wide-field, wide-band surveying instrument of the entire northern hemisphere at low frequencies, though it's important to note that LOFAR effectively enables a host of other scientific goals such as real-time monitoring of the northern sky at low frequencies \citep{2016JAI.....541008P}.

While these instruments take decades to envision and build, arguably most of the technological difficulty follows after the instrument comes online as scientists come to understand the instruments.
For LOFAR, despite coming online in 2012, after six years of construction, it was not until 2016 that a suitable direction dependent (DD) calibrated image was produced for its high-band antennas (HBA; 115--189MHz) \citep{2016ApJS..223....2V}, and though there has been much progress there is still currently no DD calibrated image produced for its low-band antennas (LBA; 42--66MHz). 
Importantly, despite many years in developing these initial DD calibration and imaging algorithms, they were not suitable for LOFAR's primary survey objective.
Thus, there has been continuous work aimed at improving them.

The main cause of DD effects in radio interferometry is the ionosphere.
The ionosphere is a turbulent, low-density, multi-specie ion plasma encompassing the Earth, driven mainly by solar extreme ultra violet radiation \citep{1995isp..book.....K}, changing on the timescale of tens of seconds \citep{1952JATP....2..141P}.
The free electron density (FED) of the ionosphere gives rise to a spatially and temporally varying refractive index at radio wavelengths.
This causes weak-scattering of electric fields passing through the ionosphere \citep{1956RPPh...19..188R}, becoming more severe at lower frequencies, and resulting in the scattering of radio radiation onto radio interferometers.
Despite the weak-scattering conditions, the integral effect of repeated scattering through the thick ionosphere can cause dispersive phase variations of well over 1 radian on the ground \citep{1951RSPSA.209...81H, 1952RSPSA.214..494H}.

Wide-field radio arrays, such as LOFAR, are particularly susceptible to the direction dependent (DD) effects of the ionosphere \citep{2009AJ....138..439C}, since the field of view is many isoplanatic-patches \citep{1955RSPTA.247..369F} wherein the instantaneous scattering properties of the ionosphere changes significantly over the field of view. 
In order to achieve the science goals in the wide-field regime, a DD calibration strategy must be adopted \citep{2005ASPC..345..350C}.
Many DD approaches have been developed ranging from field-based calibration \citep{2004SPIE.5489..180C}, expectation-maximisation \citep{2011MNRAS.414.1656K} and facet based approaches \citep{2009A&A...501.1185I, 2016ApJS..223....2V, 2018A&A...611A..87T}.
%TODO: maybe include MWA's field-based citation

Common to all of these DD techniques is the requirement of enough signal-to-noise on short enough time intervals where the ionosphere can be considered fixed.
Therefore, bright compact sources are required throughout the field of view, which act as in-field calibrators.
The fundamental assumption is that the distance between calibrators is less than the isoplanatic-patch size, which is about $1^\circ$ for nominal ionospheric conditions.
However, this quantity is dynamic over the course of an observation, and this assumption can be broken at times.

With the advent of the killMS DD calibration algorithm \citep{2014arXiv1410.8706T, 2015MNRAS.449.2668S}, and the DDFacet DD imager \citep{2018A&A...611A..87T}, the LOFAR Two-Metre Sky Survey \citep[LoTSS;][]{2017A&A...598A.104S} became possible and LoTSS first data release (DR1) has become available \citep{2019A&A...622A...1S}.
While LoTSS-DR1 provides an excellent median sensitivity of $S_{\rm 144MHz} = 71~\mu\mathrm{Jy}\,\mathrm{beam}^{-1}$ and point-source completeness of 90\%, there are still significant DD effects in the images between the in-field calibrators.
%8\% of fields completely failed due to calibration artefacts, which appears to be partly due to ionospheric effects.
% On top of this, of the successful observations, 
Improving the DD calibration for LoTSS is therefore a priority.

Data release two (DR2) is preliminarily described in \citet{2019A&A...622A...1S}, and will be fully described in (Tasse et al. in prep.).
Since we have internal access to this archive, we will make reference to LoTSS-DR2.

% For time-domain radio astronomy the ionosphere is a significant nuisance depending on the observing frequency.
% For AARTFAAC at 57MHz the amplitude variations due to the ionosphere are the main source of detection uncertainty on the shortest timescales \citep{2019MNRAS.482.2502K}.
% For the MWA at 154MHz the time-domain effects of the ionosphere are not significant \citep{2015MNRAS.453.2731L}.
% For the EoR  \citep{2013ExA....36..235M}.

The aim of this work is to present a solution to this limitation by probabilistically inferring the DD calibrations for all sources in a field of view given information of the DD calibrations for a sparse set of bright calibrators.
Importantly, this approach can be treated as an add-on to the existing LoTSS calibration and imaging pipeline.

We arrange this paper as follows.
In Section~\ref{sec:ddtec} we formally define the problem of DD calibration of ionospheric effects via inference of doubly differential total electron content.
In Section~\ref{sec:method} we describe our procedure used to perform DD calibration and imaging of a randomly chosen LoTSS-DR2 observation using doubly differential total electron content screens.
In Section~\ref{sec:results} we compare our image with the LoTSS-DR2 archival image.

\section{Doubly differential phase screens}
\label{sec:ddtec}
%TODO: consider replace calibration and imaging with CI since it often appears in the text

In radio interferometry the observable quantity is the collection of spatial auto-correlations of the electric field, which are called the visibilities.
The relation between the electric field intensity and the visibilities is given by the Radio Interferometry Measurement Equation \citep[RIME;][]{1996A&AS..117..137H},
\begin{align}
    \bs{V}(\x, \x') = \sum_{\k \in \mathcal{K}} & \left(\B(\x, \k)\J(\x,\k) \otimes \J^\dagger(\x',\k) \B^\dagger(\x', \k)\right) \langle \E(\k) \otimes \E^\dagger (\k)\rangle \notag\\
    &\times h(\x, \k) h^*(\x', \k),\label{eq:rime}
\end{align}
where we've left out frequency dependence for visual clarity, though for future reference we let there be $N_{\rm freq}$ sub-bands of the bandwidth.

In order to understand Eq.~\ref{eq:rime}, let the celestial radio sky be formed of a countably infinite number of discrete point sources with directions in $\mathcal{K} = \{\k_i \in \mathbb{S}^2 \mid i=1..N_{\rm dir}\}$.
The propagation of a monochromatic polarised celestial electric field $\E(\k)$, can be described by the phenomenological Jones algebra \citep{1941JOSA...31..488J}.
Letting the locations of antennas be $\mathcal{X} = \{\x_i \in \mathbb{R}^3 \mid i=1..N_{\rm ant}\}$, then $\E(\x,\k) = h(\x, \k)\J(\x, \k)\E(\k)$ denotes propagation of $\E(\k)$ to $\x$ via the optical pathway indexed by $(\x, \k)$, where $h(\x,\k)$ is the Huygens-Fresnel propagator in a vacuum.
Each antenna has a known instrumental optical transfer function which gives rise to another Jones matrix known as the beam, $\B(\x, \k)$.
The total electric field measured by an antenna at $\x$ is then $\E(\x) = \sum_{\k \in \mathcal{K}} h(\x,\k)\B(\x,\k)\J(\x, \k)\E(\k)$ and Eq.~\ref{eq:rime} follows from considering the auto-correlation of the electric field between antennas and imposing the physical assumption that the celestial radiation is incoherent.

Some words on terminology; despite the fact that $h$, $\B$, and $\J$ are all Jones operators, we will only refer to $\J$ as the Jones matrices going forward, and will index them with a tuple of antenna location and direction $(\x,\k)$ -- which corresponds to an optical pathway.
When a Jones matrix is a scalar times identity, the Jones matrix commutes as a scalar, and we will simply use scalar notation.
This is the case, e.g. when the Jones matrices are polarisation independent.
It is also common terminology to call the sky-brightness distribution, $\I(\k) \triangleq \langle \E(\k)\otimes\E^\dagger (\k)\rangle$, the image or sky-model, and values of the sky-brightness distribution as components.

In radio astronomy we measure the visibilities and invert the RIME iteratively for the sky-model and Jones matrices.
The process is broken into two steps called calibration and imaging.
In the calibration step the current estimate of the sky-model is held constant and the Jones matrices are inferred.
In the imaging step the current estimate of the Jones matrices are held constant and the sky-model is inferred.
In both the imaging and calibration steps the algorithm performing the inference can be either direction independent (DI) or direction dependent (DD), which stems from the type of approximation made on the RIME when modelling the visibilities.
Namely, the DI assumption states that the Jones matrices are the same for all directions, and the DD assumption states that the Jones matrices have some form of directional dependence.

In the current work we focus entirely on the Stokes-I, i.e. polarisation independent, calibration and imaging program used by the LoTSS pipeline.
The LoTSS calibration and imaging pipeline is broken up into a DI calibration and imaging step followed by a DD calibration and imaging step.

As can be seen from Eq.~\ref{eq:rime}, each optical pathway $(\x, \k)$ has its own unique Jones matrix $\J(\x, \k)$ describing the propagation of radiation.
In total, this results in $2 N_{\rm dir} N_{\rm ant}$ degrees of freedom for $\frac{1}{2}N_{\rm ant}(N_{\rm ant}-1)$ observables.
In practice, during DD calibration and imaging this is never assumed for three main reasons. 
Firstly, the computational memory requirements of giving every direction a Jones matrix are prohibitive, secondly, since $N_{\rm dir} \gg N_{\rm ant}$ it would allow too many degrees of freedom per observable which leads to ill-conditioning, and thirdly, most sky-model components are too faint to infer a Jones matrix on the timescales that the ionosphere can be considered fixed.

The LoTSS calibration and imaging program manages to handle the computational memory requirement and ill-conditioning via a clever sparsification of the optimisation Jacobian and by invoking an extended Kalman filter respectively; see \citet{2014arXiv1410.8706T, 2014A&A...566A.127T, 2015MNRAS.449.2668S} for details.
However, there is no way to prevent the third issue of too little signal on required timescales, and they are limited to tessellating the field of view into approximately $N_{\rm cal} \approx 45$ facets, defined by a set of calibrator directions, $\mathcal{K}_{\rm cal} \subset \mathcal{K}$.
The result of DI and DD calibration and imaging for LoTSS is therefore a set of $N_{\rm ant}$ DI Jones matrices, and a set of $N_{\rm cal} N_{\rm ant}$ DD Jones matrices.

Let us consider the functional form of the final Jones matrices following this two part calibration and imaging program.
For each $(\x, \k) \in \mathcal{X}\times\mathcal{K}_{\rm cal}$ we have the associated calibration Jones matrix,
\begin{align}
    \J^{\rm cal}(\x, \k) \triangleq& \J^{\rm DD}(\x, \k) \J^{\rm DI}(\x, \k')\\
    =& J^{\rm DD}(\x, \k) J^{\rm DI}(\x, \k') e^{\iota (\Delta_0\phi^{\rm DD}(\x, \k) + \Delta_0\phi^{\rm DI}(\x, \k'))},\label{eq:jones_cal_a}
\end{align}
where notation $t \triangleq m$ means definition by equality, and can be read as `$t$ is equal by definition to $m$'.
An effective direction can intuitively be defined for the DI Jones matrices as the direction, $\k'$, that minimises the instantaneous image domain dispersive phase error effects in the DI image.
Clearly, this leads to problems for a wide-field instrument, where the actual Jones matrices can vary considerably over the field of view.
In fact, it's fairly simple to contrive examples where there is no well-defined effective direction.
Nonetheless, we will use this terminology and come back to address the associated problems shortly.

From the Wiener-Khinchin theorem it follows that the visibilities are independent of the electric field phase, therefore a choice is made to spatially phase reference the Jones matrices to the location $\x_0$ of a reference antenna. 
Doing so defines the differential phase, 
\begin{align}
    \Delta_0 \phi(\x, \k) \triangleq \phi(\x, \k) - \phi(\x_0, \k).
\end{align}

Now, consider a set of observed visibilities, which are perfectly described by Eq.~\ref{eq:rime}, where we are neglecting time, frequency, or baseline averaging.
Then, there must be a set of true Jones matrices, $\mathcal{J}^{\rm true} = \{\J^{\rm true}(\x, \k) \mid (\x, \k) \in \mathcal{X}\times\mathcal{K}\}$, which give rise to the observed visibilities.
We shall assume such a set of Jones matrices is unique.
Because the Jones matrices are assumed to be scalars, we can write them as $\J^{\rm true}(\x, \k) \triangleq J^{\rm true}(\x, \k) e^{\iota \Delta_0\phi^{\rm true}(\x, \k)}$.

Next, consider the process of inferring the true Jones scalars from the observed visibilities using piece-wise constant calibration Jones scalars, $\J^{\rm cal}(\x, \k)$, as is done in killMS.
To do so, we define the operation $\floor{\k} = \argmax_{\k' \in \mathcal{K}_{\rm cal}} \k\cdot\k'$ as the closest calibration direction to a given direction.
In general, it shall always be possible to write the true Jones scalars, $\J^{\rm true}(\x, \k)$, in terms of the resulting calibration Jones scalars $\J^{\rm cal}(\x, \k)$ as,
\begin{align}
    \J^{\rm true}(\x, \k) \triangleq& \alpha(\x, \k) e^{\iota \Delta_0\beta(\x, \k)} \J^{\rm cal}(\x, \floor{\k}),\label{eq:J_true_a}
\end{align}
where $\alpha(\x, \k)$ and $\Delta_0\beta(\x, \k)$ correspond to the correction factors that would need to be applied to each calibration Jones scalar to make it equal to the true Jones scalar along that optical pathway.
Let us assume the situation where there is enough information in the observed visibilities to strongly constrain the calibration Jones scalars, and that we are able to point-wise solve for a global optimum -- which is unique by our above assumption.
Then, if the distance between calibrators is much less than the isoplanatic-patch size \citep{1955RSPTA.247..369F}, we have that $\alpha \approx 1$ and $\Delta_0\beta \approx 0$.

The substance of this paper is aimed at inferring the function $\Delta_0\beta$ for all optical pathways (we'll pay minor attention to $\alpha$ as well).
Due to the dynamic nature of the ionosphere the calibration facets are not always isoplanatic ($\k \not\approx \floor{\k}$ in the appropriate sense defined in \citet{1955RSPTA.247..369F}) and consequently $\Delta_0\beta \not\approx 0$ sometimes.

Let us now restrict our attention to the calibration Jones matrices in the direction of the calibrators and consider the value of $\Delta_0\beta(\x, \k)$ for $(\x,\k)\in  \mathcal{X}\times\mathcal{K}_{\rm cal}$.
In the event that $\Delta_0\beta(\x, \k) \approx 0$, then from Eqs.~\ref{eq:J_true_a} and \ref{eq:jones_cal_a} it follows that the DD differential phases are,
\begin{align}
    \Delta_0\phi^{\rm DD}(\x, \k)  =  \Delta_0\phi^{\rm true}(\x, \k) - \Delta_0\phi^{\rm DI}(\x, \k').\label{eq:jones_dd_a}
\end{align}
This states that in the directions of the calibrators, that the DD calibration Jones phases are equal to the difference between the true Jones phases and DI Jones phases.
However, in a typical application of facet-based calibration, we can easily see that this statement is usually false.
It basically comes down to the fact that the designation of DI and DD for the Jones matrices inferred in calibration and imaging is a misnomer.

Firstly, specifying that the DI Jones matrices have a single effective direction is an invalid statement.
The effective direction, as defined above, minimises the instantaneous image domain dispersive phase error effects, and this depends largely on the ionosphere, geometry of the optical pathways, and the sky-brightness distribution.
The ionosphere changes in time, acting as a time-varying scattering layer in the optical system, therefore the effective direction is also dynamic.
The effective direction is heavily weighted towards the brightest sources in the field of view, and since longer baselines `see' less flux, the effective direction is different for each antenna.
To a lesser extent the sensitivity drop-off at lower frequencies will also introduces a effective direction that changes with sub-band frequency.

Secondly, despite the fact that DI calibration should remove all DI components, such as station clocks, in practice there are always remnant DI components in the DD Jones matrices.

It shall be desirable to assume Eq.~\ref{eq:jones_dd_a} holds in going forward, as we shall interpret this to mean that we can make $\Delta_0 \beta(\x, \k) \approx 0$ everywhere through an appropriate choice of $\Delta_0 \phi^{\rm DD} (\x, \k)$.
Therefore we consider how to modify DD calibration such that Eq.~\ref{eq:jones_dd_a} holds.

We address the first issue above, of ill-defined effective direction, by suitably forcing the observed visibilities to be well-approximated by isolated calibrators.
This is applied and discussed further in Section~\ref{sec:subtract_and_solve}.

To correct the second issue, to exactly separate all DI components from the DD Jones matrices, we utilise a trick.
Notice in Eq.~\ref{eq:rime} that we can factor all commutative DI factors out in front of the summation.
As we show in Appendix~\ref{app:ddtec}, this is equivalent to directionally phase referencing the Jones scalars, which we denote as doubly differential phase,
\begin{align}
    \Delta_0^2 \phi \triangleq \Delta_0\phi(\x, \k) - \Delta_0 \phi (\x, \k_0).
\end{align}
Moreover, we show that doubly differential phase is guaranteed to have no DI components.
This trick is used and discussed further in Section~\ref{sec:vi_ddtec}.

After applying the above two fixes, Eq.~\ref{eq:jones_dd_a} then takes the form of a doubly differential phase, $\Delta_0^2 \phi^{\rm DD}(\x, \k)$. 
These not only have no DI components, but also have a well-defined reference direction.
Both attributes are vital for our modelling method.

Finally, this paper is concerned with performing DD calibration by inferring $\Delta_0^2 \phi^{\rm DD}(\x, \k)$ for a set of isolated calibrators followed by interpolation of these values to a multitude of optical pathways, which we shall call a screen.
We shall assume that the DD Jones scalars are dominated by ionospheric dispersive phase effects and slowly changing beam errors.
We neglect the ionospheric amplitude effects in this description since the phase effects are far more dominant.
When the frequency of radiation is far above the ionospheric plasma frequency, then the ionospheric phases effects are given by a dispersive phase.
The dispersive phase retards the wavefront without distorting it.
The doubly differential ionospheric dispersive phase is given by,
\begin{align}
    \Delta_0^2 \phi^{\rm ion}(\x, \k) = \frac{\kappa}{\nu} \Delta_0^2 \tau(\x, \k),\label{eq:ddtec_a}
\end{align}
where $\Delta_0^2 \tau(\x, \k)$ is doubly differential total electron content (DDTEC), and $\kappa = \frac{-q^2}{4 \pi \epsilon_0 m c}$.

\section{Methods}
\label{sec:method}

In this paper we push beyond the state-of-the-art LoTSS DD calibration and imaging by using a physics-informed probabilistic model to infer DDTEC screens. 
We perform our screen-based DD calibration and imaging on a randomly selected 8-hour observation from the LoTSS-DR2 archive.
The selected observation, LoTSS-DR2 pointing P126+65, was observed between 23:12:14 on 8 December, 2016 and 07:28:34 on 9 December, 2016, during solar cycle 24\footnote{On the 8 December, 2016 the total number of sunspots was $14\pm1.2$ and on 9 December, 2016 it was $20\pm5.7$. The average daily sunspot count in December 2016 was 18.}.
The ionosphere activity during this observation was mixed, with the first half of the observation `wild' -- DDTEC exceeding 100mTECU -- and the last half `calm'.

\begin{figure}
    \centering
    \includegraphics[width=\columnwidth]{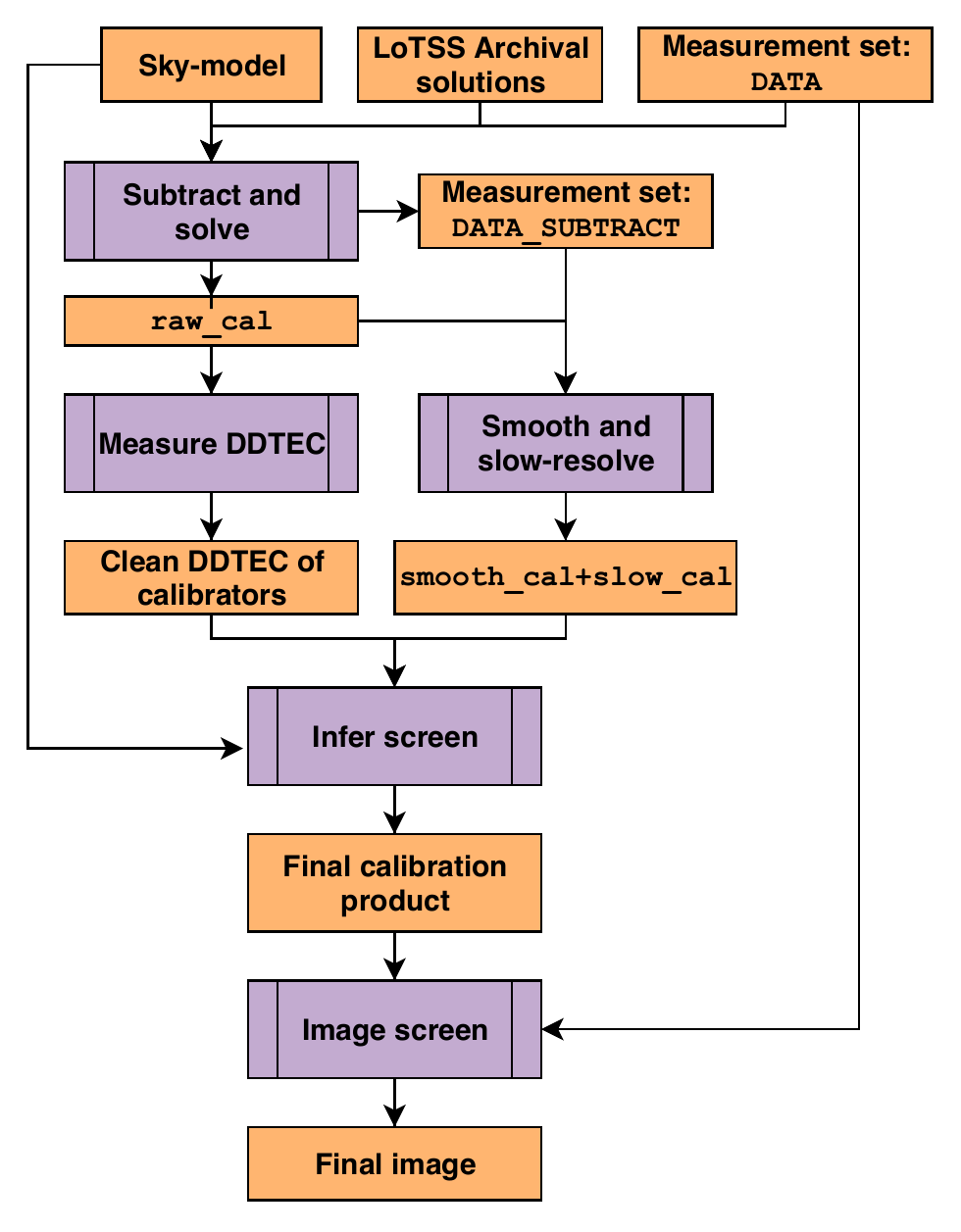}
    \caption{Flow diagram overview of our DD calibration and imaging program.}
    \label{fig:overview}
\end{figure}

We first provide an overview of the our DD calibration and imaging procedure, before providing a deeper explanation of each step.
A summary of our DD calibration and imaging program can be seen in \textbf{Figure}~\ref{fig:overview} and is as follows:
\begin{enumerate}
    \item \textit{Subtract and solve step}. Subtract a good model of the sky, except approximately 35 bright calibrators (peak flux > 0.3~$\mathrm{Jy}\,\mathrm{beam}^{-1}$). Solve against isolated calibrators.
    \item \textit{Smooth and slow-resolve}. Smooth the Jones scalars, and resolve on a long time scale to simultaneously reduce the degrees of freedom and solve the holes problem \citep[for a description of the problem see][]{2019A&A...622A...1S}.
    \item \textit{Measure DDTEC}. Infer DDTEC from the Jones scalars in the directions of the bright calibrators.
    \item \textit{Infer DDTEC screen}. Apply our DDTEC model, to infer DDTEC for all intermediate brightness optical pathways (>0.01~$\mathrm{Jy}\,\mathrm{beam}^{-1}$), forming a screen over the field of view.
    \item \textit{Image screen}. Concatenate the smoothed solutions with the DDTEC screen. Image the original visibilities with this set of solutions.
\end{enumerate}

\subsection{Subtract and solve}
\label{sec:subtract_and_solve}

\begin{figure}
    \centering
    \includegraphics[width=\columnwidth]{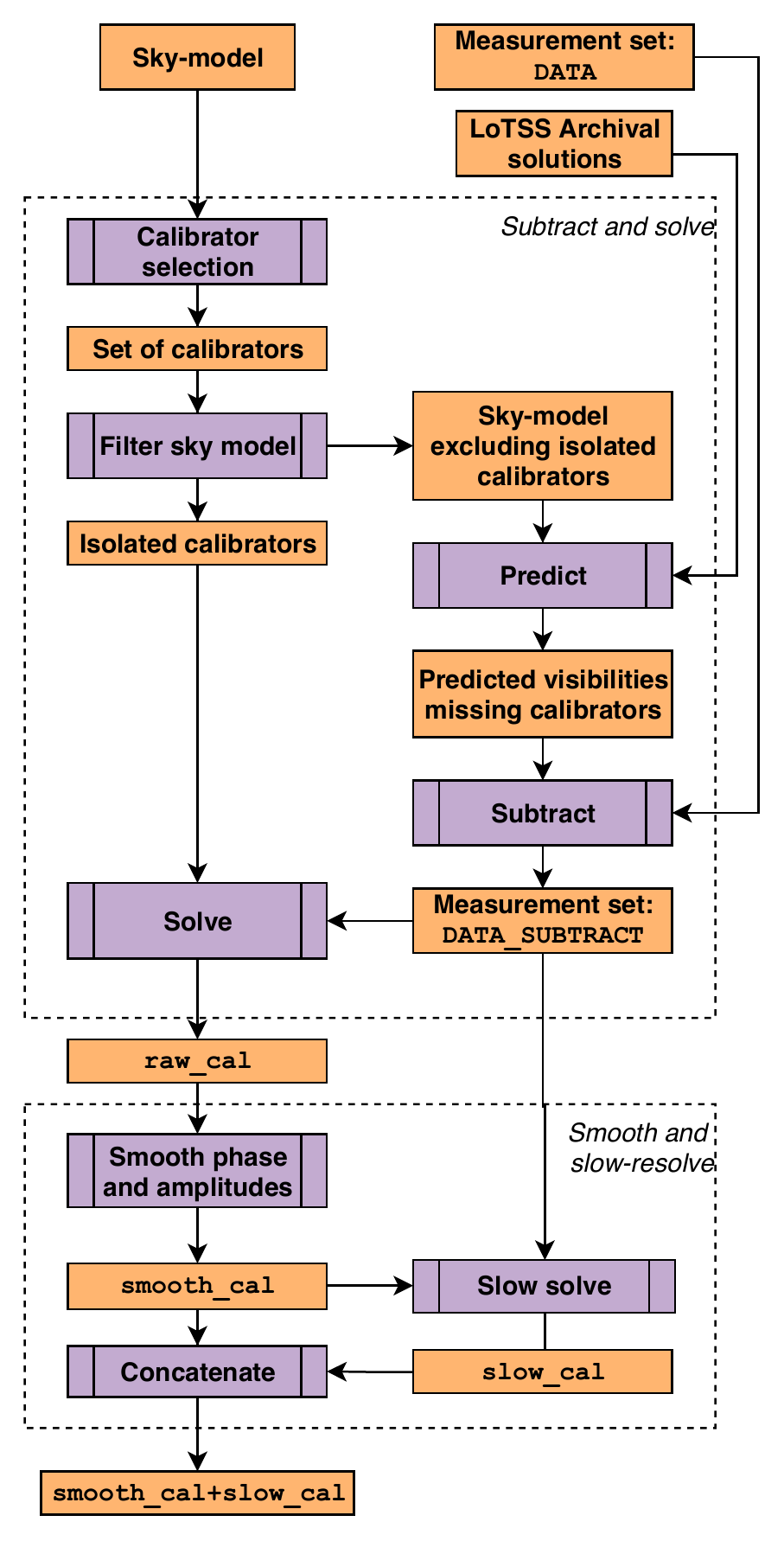}
    \caption{A flow chart of the \textit{subtract and solve} and \textit{smooth and slow-resolve} steps. Together they represent a DD calibration program similar to the LoTSS-DR1 and DR2 pipelines.}
    \label{fig:first_two_steps}
\end{figure}

In this step we isolate a set of calibrators by subtracting the complementary sky-model and then perform a solve. 
The process is visualised in the correspondingly labelled process box in \textbf{Figure}~\ref{fig:first_two_steps}.

\begin{figure}
    \centering
    \includegraphics[width=\columnwidth]{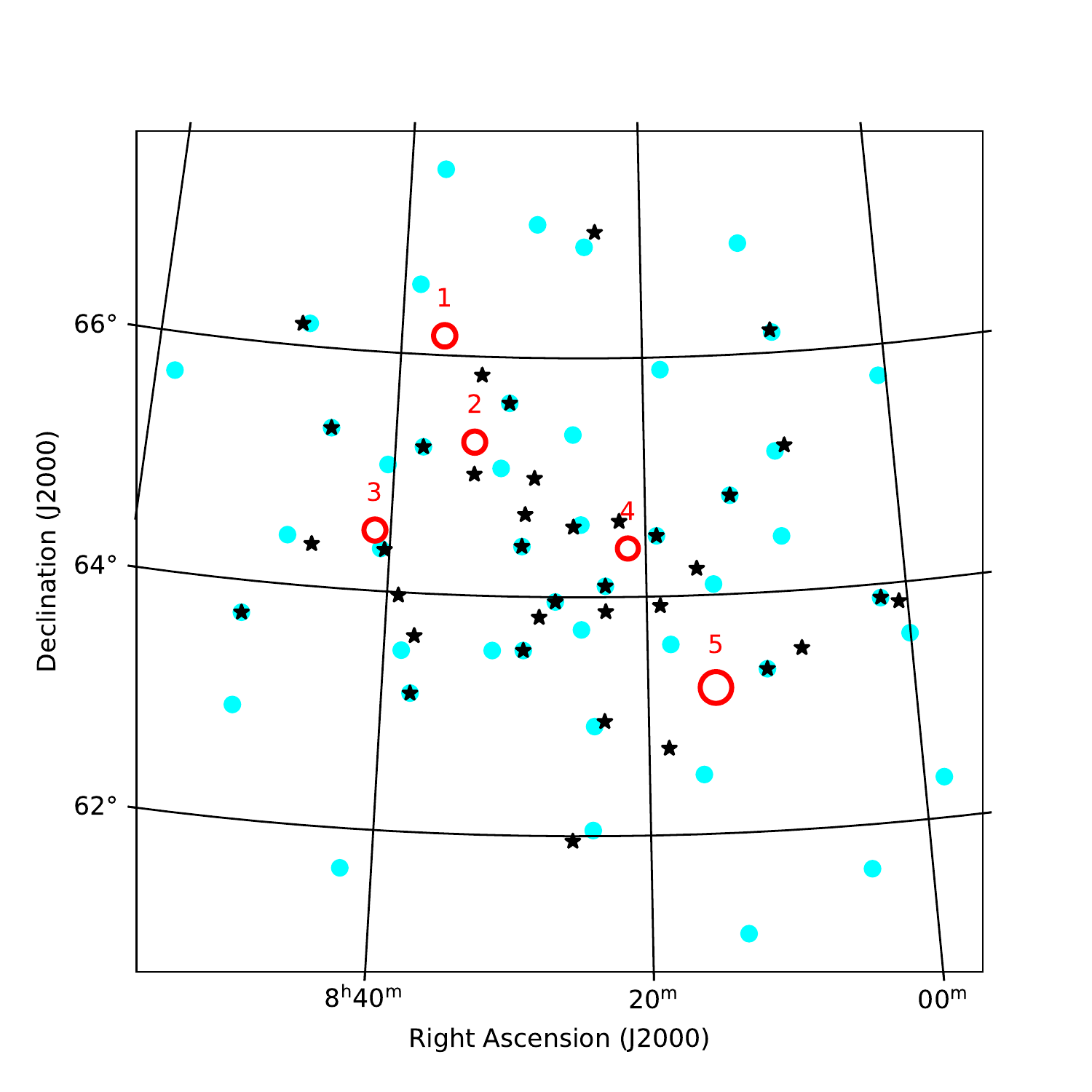}
    \caption{Layout of our selected calibrators (black stars) compared to the LoTSS-DR2 archival calibrator layout (cyan circles).
    The calibrators define a facet tessellation of the field of view.
    The red circles indicate the regions of the cutouts in \textbf{Figure}~\ref{fig:image_panel}.
    }
    \label{fig:field_regions}
\end{figure}

We begin by selecting a set of bright calibrators (peak flux>0.3~$\mathrm{Jy}\,\mathrm{beam}^{-1}$).
The goal is to cover the field of view as uniformly as possible, while not selecting too many calibrators, which can result in ill-conditioning of the system of equations that must be solved.
In practice, we usually do not have enough bright calibrators to worry about ill-conditioning.
Our selection criteria resulted in 36 calibrators whose layout is shown in \textbf{Figure}~\ref{fig:field_regions} alongside the 45 calibrators selected by the LoTSS-DR2 pipeline.

\begin{figure}
    \centering
    \includegraphics[width=\columnwidth]{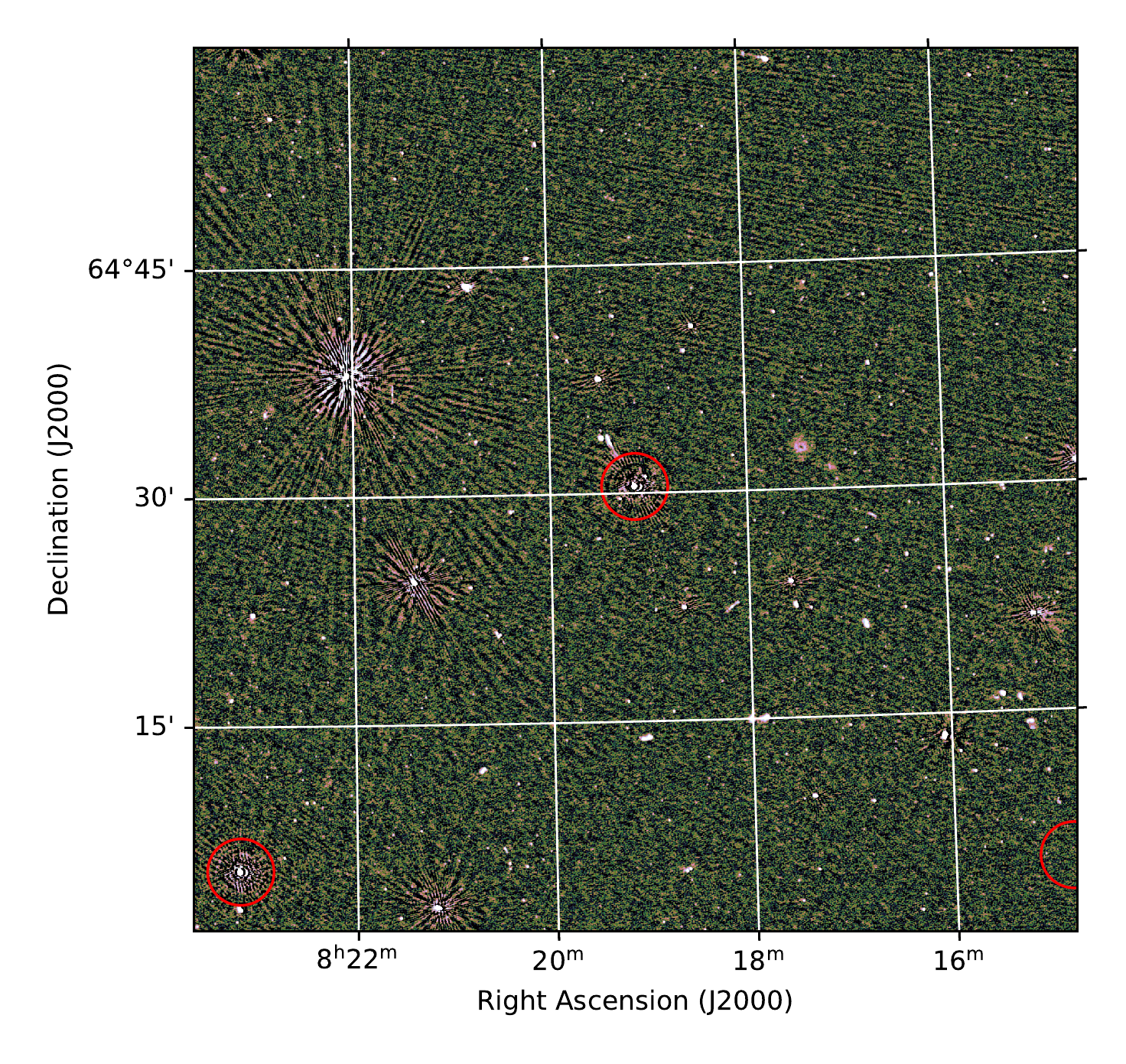}
    \caption{Example of average effective direction in a facet.
    The red circles indicate calibrators. The central calibrator has a peak flux of 0.91~$\mathrm{Jy}\,\mathrm{beam}^{-1}$, while the next brightest non-calibrator source, approximately 
    $16\arcmin$ to the left, is 0.12~$\mathrm{Jy}\,\mathrm{beam}^{-1}$. 
    Within the facet of the calibrator, the Jones matrices are mostly optimised in a direction of the calibrator, however this effective direction changes over time, baseline, and frequency.}
    \label{fig:eff_dir}
\end{figure}

As stated in Section~\ref{sec:ddtec}, for our method to work it is vital that the directions of the calibrators are well-defined.
The Jones matrix for each calibrator can be considered as a DI Jones matrix within the calibrator facet, therefore the notion of effective direction, as the direction that minimises instantaneous image domain artefacts, applies within a facet as well.
Intuitively, the effective direction is located near the brightest sources in a facet.
For example, in \textbf{Figure}~\ref{fig:eff_dir} the central source is by far brighter than all nearby sources, consequently the dispersive errors in that direction are minimised.
This defines an average effective direction.

Since the effective direction changes over time, baseline, and frequency, it follows that DD modelling will be systematically biased by the unknown direction of the Jones matrices if they are based on these ill-defined effective directions.
We circumvent this issue by first subtracting all sources from the visibilities, except for the bright calibrators.
This requires having a good enough initial sky-model.
In our case we use the sky-model from the LoTSS-DR2 archive.
We mask a $120\arcsec$ disk around each calibrator, ensuring that all artefacts for the calibrators are included in the mask.
We predict the visibilities associated with the remainder of the sky-model, pre-applying the solutions from the LoTSS-DR2 archive.
We then subtract these predicted visibilities from the raw visibilities in the \texttt{DATA} column of the measurement set, placing the result in a \texttt{DATA\_SUBTRACT} column of the measurement set.
Once subtracted, the isolated calibrator sources will provide well-defined directions, which is crucial for Eq.~\ref{eq:jones_dd_a} to hold.

We note that, any unmodelled flux missing from the sky-model will still reside in the visibilities leading to a systematic calibration bias when inverting the RIME.
However, they will have been missing from the model due to their inherent faintness, therefore they should have minimal effect on the effective directions inside the facets of the isolated calibrators.

It is possible that a bright extended source be included in in the set of calibrators.
In this case, the model of the extended source will be truncated at a radius of $120\arcsec$ from the peak.
Because the sky-model is necessarily incomplete, there will be residual flux of the extended source left over.
As above, due to the faintness of these residual components, they will have negligible effects on the resulting effective direction of the calibrator.
Another problem of choosing extended sources as calibrators is that of convergence due to their resolved nature.
Extended calibrators can lead to divergent calibration Jones scalars, which is a problem that must also be handled by any calibration and imaging program.
In our case, the sky-model comes from the LoTSS-DR2 archive which provides a carefully self-calibrated model at the same frequency and resolution, and we expect to suffer less from divergent solutions.

% Due to LOFAR's relatively high sensitivity outside its primary lobe, unmodelled bright sources outside the primary beam can significantly contributed to the observed visibilities.
% By construction, unmodelled sources within the primary beam are very faint, and do not significantly contribute to the observed visibilities.

We perform a solve on \texttt{DATA\_SUBTRACT} against the sky-model components of the isolated bright calibrators.
We call these solutions \texttt{raw\_cal} for future reference.

There are several non-ionospheric systematics that will be in \texttt{raw\_cal}.
Firstly, beam errors are a systematic that come from performing calibration with an incomplete beam model, as well as with an aperture array that contains dead or disturbed array tiles.
The beam model can be improved by better physical modelling of the aperture array, which requires a complex solutions to Maxwell's equations.
The second issue is a transient effect, since tiles can go dead over time and the dielectric properties of the antennae environment can change with temperature and air moisture content.
It will be important in the following to account for beam errors. 
Beam errors \textit{a priori} should change slowly over the course of an observation.

Another systematic that will find it's way into the \texttt{raw\_cal} are unmodelled sources outside the image, in LOFAR's relatively high-sensitivity side-lobes.
The effect of such unmodelled sources on calibration is still being understood.
As noted in \citet{2019A&A...622A...1S}, such sources can be absorbed by extra degrees of freedom during calibration.
Since they are not within the imaged region, this flux absorption can go unnoticed.
It shall therefore be important in the following to limit the degrees of freedom of the calibration Jones scalars.

\subsection{Smooth and slow-resolve}
\label{sec:smooth}

There are enough degrees of freedom in a DD solve that unmodelled flux can be absorbed \citep{2019A&A...622A...1S}, therefore naively applying \texttt{raw\_cal} and imaging would be detrimental to source completeness.
\citet{2019A&A...622A...1S} found that a post-processing step of Jones scalar phase and amplitude smoothing removed enough degrees of smoothing to alleviate this problem.
However, they then discovered that such a smoothing operation introduces negative halos around sources, the systematic origin of which has yet to be determined.
A second solve on a much longer time scale alleviates the negative halo problem.
The process is visualised in the correspondingly labelled process box in \textbf{Figure}~\ref{fig:first_two_steps}.

The LoTSS DD calibration and imaging pipeline performs Jones scalar phase smoothing by fitting a TEC-like and constant-in-frequency functional form to the Jones phases using maximum likelihood estimation.
This results in two parameters per 24 observables.
We alter the smoothing method by fitting a three parameter model: a TEC-like term, a constant-in-frequency, and a clock-like functional form.
Our choice results in three parameters per 24 observables.
We add the extra term because a small ($\lesssim 1~\mathrm{ns}$) remnant clock-like term appears to exist in the DD Jones phases.

Weak scattering in the ionosphere causes a well-known DD amplitude effect, however the amplitudes of the Jones scalars in \texttt{raw\_cal} are more complicated, with possible contamination from an improperly modelled beam and unmodelled flux.
Therefore, we do not consider modelling amplitude in this work, however we do impose the prior belief that beam error amplitudes should change slowly over time and frequency as in \citet{2019A&A...622A...1S}.
To impose this prior, we simply smooth the log-amplitudes of \texttt{raw\_cal} with a two-dimensional median filter in time and frequency with window size (15~min., 3.91~MHz).

Scintillation can also significantly effect the amplitudes of the Jones scalars.
In the case where the scintillation timescale is shorter than the integration timescale, the observed visibilities can decohere.
In such a case, the visibilities should be flagged.
DDFacet contains an adaptive visibility weighting mechanism inspired by the optical method of "lucky imaging" that down-weights noisy visibilties \citep{2018A&A...615A..66B}, which effectively handles fast scintillation.
In the case that scintillation occurs on a longer timescale, the effect is characterised by amplitude jumps for groups of nearby antennae within the Fresnel zone.
These effects are predominately removed during the DI calibration and are not considered further.
This implies that we do not consider DD scintillation.

A surprising result of the smoothing procedure is that it introduces negative halos around bright sources, which destroys the integrity of the image.
In the LoTSS pipeline, despite much effort to understand the fundamental cause of the problem, a temporary solution is to pre-apply the smoothed solutions and resolve against the same sky-model on a long time scale.
We perform this resolve with a solution interval of 45~minutes call these solutions \texttt{slow\_cal}.
We form the concatenation of \texttt{smooth\_cal} and \texttt{slow\_cal} and call it \texttt{smooth\_cal+slow\_cal}.

\subsection{Measure DDTEC}
\label{sec:vi_ddtec}

Next, we extract precise DDTEC measurements from \texttt{raw\_cal}, which will later be used to infer a DDTEC screen.
While it would be possible to model a DDTEC screen directly on Jones scalars, the required inference algorithm would be extremely expensive to account for the high-dimensional, complicated, multi-modal posterior distribution that occurs due to phase wrapping.
Therefore, in this section we describe how we use the $\nu^{-1}$ dependence of dispersive phase (Eq.~\ref{eq:ddtec_a}) to measure DDTEC\footnote{We use the term `measure DDTEC' purely for distinction between the inferred DDTEC screen discussed in Section~\ref{sec:infer_screen}. Also, from a philosophical perspective, a measurement is a single realisation of inference.}.
The process is visualised in the correspondingly labelled box of \textbf{Figure}~\ref{fig:measure_ddtec_and_infer}.

First, we directionally reference the Jones phases in \texttt{raw\_cal} by subtracting the phase of the brightest calibrator -- the reference calibrator -- from the phases of all other directions.
We can assume that all remnant DI effects are removed from the resulting doubly differential phases (see Appendix~\ref{app:ddtec}).
We then assume that these Jones scalars phases are completely described by DDTEC and the smoothed amplitudes coming from beam errors. 

\begin{figure}
    \centering
    \includegraphics[width=\columnwidth]{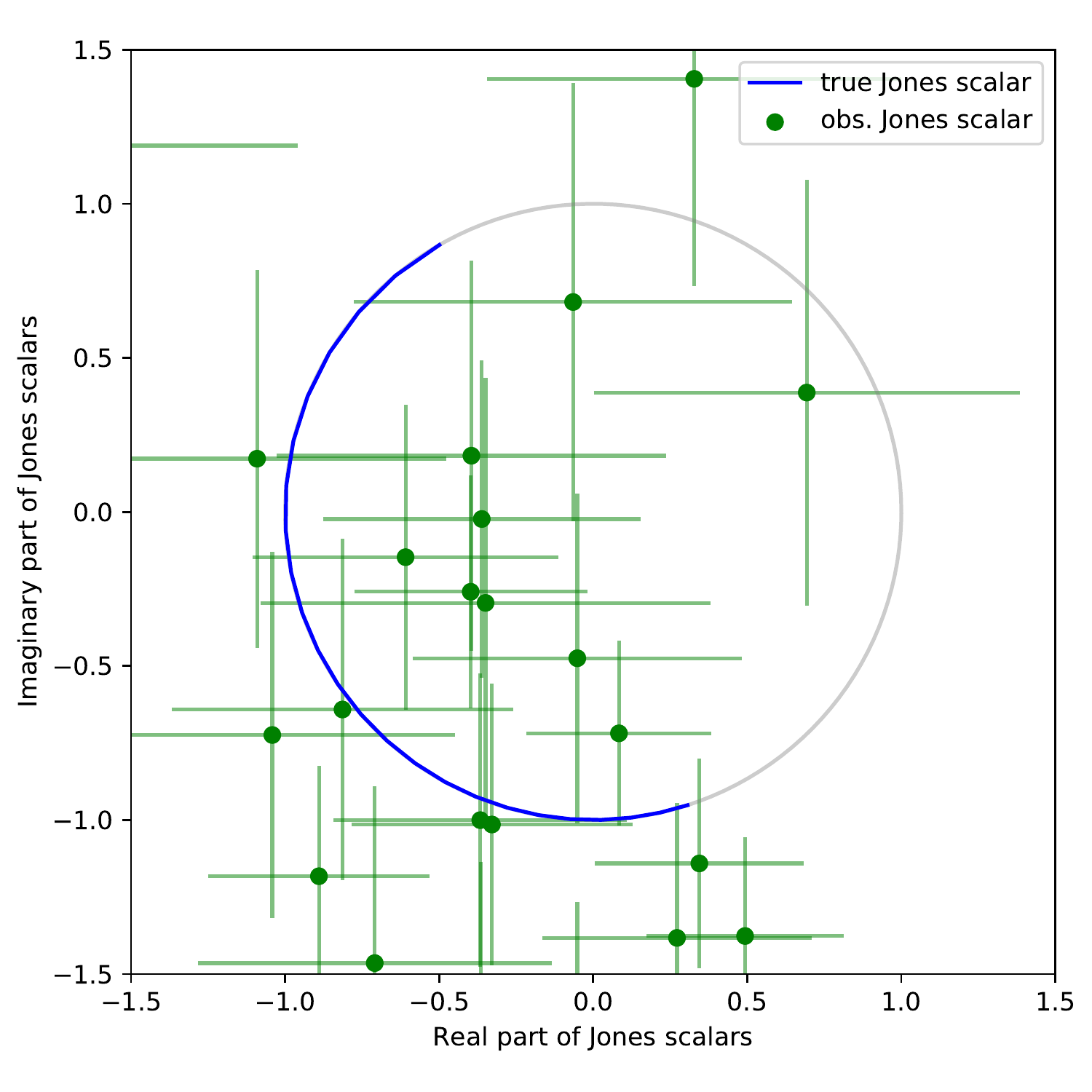}
    \caption{Simulated real and imaginary components of noise-dominated Jones scalars and observational uncertainties. The simulated ground truth DDTEC is 150~mTECU and the observational uncertainties are frequency dependent and correlated, corresponding to a remote station for a calibrator at the cutoff peak flux limit of 0.3~$\mathrm{Jy}\,\mathrm{beam}^{-1}$.}
    \label{fig:example_gains}
\end{figure}

The idea of fitting frequency $\nu^{-1}$ dependence to phase to extract a TEC-like term is not new \citep[e.g.][]{2016ApJS..223....2V}.
However, fitting ionospheric components from Jones scalars is a notoriously difficult procedure since the Jones scalars are typically noise dominated, with heteroscedastic correlated noise profiles, and there are many local minimal caused by phase-wrapping.
\textbf{Figure}~\ref{fig:example_gains} shows an example of simulated Jones scalars data for a single optical pathway with a typical observational noise profile corresponding to a remote antenna and calibrator peak flux of 0.3~$\mathrm{Jy}\,\mathrm{beam}^{-1}$.
Typically, the problem is solved using maximum-likelihood estimators and the resulting TEC-like parameters are too biased or high-variance to be used for any sort of inference.
For our purposes we require uncertainties of the measured DDTEC, which should be on the order of a few mTECU.

We formulate the problem as a hidden Markov model \citep[HHM;][]{rabiner1986}.
In this HMM the Jones scalars of an optical pathway form a sequence of observables indexed by time, and the DDTEC of the optical pathway forms a Markov chain of hidden variables.
We then apply recursive Bayesian estimation to get the posterior distribution of the DDTEC at each point in time given the entire Jones scalar sequence.
As shown in Appendix~\ref{app:rbe} this can be done in two passes over the data, using the forward and backward recursions, Equations~\ref{eq:update} and \ref{eq:backwards} respectively.

Let $\g_i \in \mathbb{C}^{N_{\rm freq}}$ be the observed complex Jones scalar vector at time step $i$, and let $\ddtec_i$ be the corresponding DDTEC.
For visual clarity, we drop the optical pathway designation, since this analysis happens per optical pathway.

We assume that the Jones scalars have Gaussian noise, described by the observational covariance matrix $\bs{\Sigma}$.
Thus, we define the likelihood of the Jones scalars as,
\begin{align}
    p(\g_i \mid \ddtec_i, \bs{\Sigma}) =& \mathcal{N}_\mathbb{C}[\g_i \mid |\tilde{\g}_i| e^{\iota \frac{\kappa}{\nu} \ddtec_i}, \bs{\Sigma}],\label{eq:vi_likelihood}
\end{align}
where $|\tilde{\g}_i|$ denotes the smoothed amplitudes, and $\mathcal{N}_\mathbb{C}$ denotes the complex Gaussian distribution.
The complex Gaussian distribution of a complex random variable is formulated as a multivariate Gaussian of the stacked real and imaginary parts.

Because we expect spatial and temporal continuity of FED, the DDTEC should also exhibit such continuity. 
Therefore, we assume that in equal time intervals the DDTEC is a L\'evy process with Gaussian steps, 
\begin{align}
p(\ddtec_{i+1} \mid \ddtec_i) = \mathcal{N}[\ddtec_{i+1} \mid \ddtec_i, \omega^2 \delta t],\label{eq:ddtec_time_coupling}
\end{align}
where $\omega^2$ is the variance of the step-size.
This is equivalent to saying the DDTEC prior is a Gaussian random walk.

Equations~\ref{eq:vi_likelihood} and \ref{eq:ddtec_time_coupling} define the HMM data likelihood and state transition distribution necessary to compute Equations~\ref{eq:update} and \ref{eq:backwards}.

The relation between DDTEC and amplitude, and the Jones scalars is non-linear, therefore Equation~\ref{eq:update} is analytically intractable.
To evaluate the forward equation, we apply variational inference, which is an approximate Bayesian method.
Variational inference proceeds by approximating a distribution with a tractable so-called variational distribution with variational parameters that must be learned from data \citep[e.g.][]{2013arXiv1309.6835H}.
A lower-bound on the Bayesian evidence is maximised for a point-wise estimate of the variational parameters, resulting in a variational distribution that closely resembles the actual distribution.

We assume that the variational forward distribution for DDTEC, at time step $i$ given all past Jones scalars, $\g_{0:i}$, is closely approximated by a Gaussian,
\begin{align}
q(\ddtec_{i}) = \mathcal{N}[\ddtec_{i} \mid m_{i}, \gamma^2_{i}],\label{eq:vi_posterior}
\end{align}
where the mean and variance is respectively $m_{i}$ and $\gamma_{i}^2$.
When this choice is made the state transition distribution is conjugate, and, in fact, the forward recursion reduces to the well-known Kalman-filter equations, and the backward recursion reduces to the Rauch-smoother equations \citep{rauch1963}.

Consider marginalisation of the state transition distribution in Equation~\ref{eq:predict}. 
Marginalising with respect to Equation~\ref{eq:vi_posterior} at time step $i-1$ we get,
\begin{align}
    p(\ddtec_i \mid \g_{0:i-1}, \bs{\Sigma}) = \mathcal{N}[\ddtec_i \mid m_{i-1}, \omega^2  \delta t + \gamma^{2}_{i-1}]. \label{eq:vi_prior}
\end{align}
This can be viewed as the propagated prior belief in DDTEC.
The forward probability density is then given by Bayes equation,
\begin{align}
    p(\ddtec_{i} \mid \g_{0:i}, \bs{\Sigma}) = \frac{p(\g_i \mid \ddtec_i, \bs{\Sigma}) p(\ddtec_i \mid \g_{0:i-1}, \bs{\Sigma})}{p(\g_i \mid \g_{0:i-1}, \bs{\Sigma})}.
\end{align}
This is the posterior probability density given all previous observations in the observable sequence.

Variational inference proceeds by minimising the Kullbeck-Leibler divergence $\mathrm{KL}[q \mid\mid p] = \int_\mathbb{R} \log{q/p}\,\mathrm{d}q$ from the variational forward distribution $q$ to the true forward distribution $p$ with respect to the variational parameters.
Equivalently, we may maximise the following evidence lower bound objective (ELBO),
\begin{align}
    \mathcal{L}_i[q,p] =& \mathbb{E}_{q(\ddtec_{i-1})}\left[\log\frac{p(\ddtec_{i} \mid \g_{0:i}, \bs{\Sigma}) p(\ddtec_i \mid \g_{0:i-1}, \bs{\Sigma})}{q(\ddtec_{i-1})}\right]\\
    =& \mathbb{E}_{q(\ddtec_{i-1})}\left[\log p(\g_i \mid \ddtec_i, \bs{\Sigma}) \right] \notag\\
    &- \mathrm{KL}[q(\ddtec_{i-1}) \mid\mid p(\ddtec_i \mid \g_{0:i-1}, \bs{\Sigma})].\label{eq:elbo}
\end{align}
Because the variational and prior distributions are both Gaussian, there is an analytic expression for the KL term\footnote{ $\mathrm{KL}\left[\mathcal{N}[\mu_1, \sigma_1^2] \mid\mid \mathcal{N}[\mu_2, \sigma_2^2]\right] = \frac{1}{2}\left(\frac{\sigma_1^2}{\sigma_2^2} + \frac{(\mu_2-\mu_1)^2}{\sigma_2^2}  - 1 + \log\frac{\sigma_2^2}{\sigma_1^2}\right)$.}.
The expectation of the likelihood term is called the variational expectation and is analytically derived in Appendix~\ref{app:var_exp}.

One of the benefits of variational inference is that it turns a problem that requires computationally expensive Markov chain Monte Carlo methods, into an inexpensive optimisation problem.
It works well, so long as the variational distribution adequately describes the true distribution.

Since we have assumed the variational forward distribution to be Gaussian, the approximated system becomes a linear dynamical system (LDS).
Several great properties follow.
Firstly, the backward equation, Equation~\ref{eq:backwards}, is reduced to the Rauch recurrence relations \citep{rauch1963}, which are easily accessible \citep[e.g.][]{shumway1982} so we do not write them down here.
The Rauch recurrence relations are analytically tractable, therefore, so long as the variational forward distribution is a reasonable approximation, they provide a computationally inexpensive improvement to DDTEC estimation.
In contrast, performing the forward filtering problem alone utilises only half of the available information in the observable sequence.

Secondly, the unknown step size variance $\omega$, and the observational uncertainty $\bs{\Sigma}$ can be estimated iteratively using the expectation-maximisation (EM) algorithm for LDS \citep{shumway1982}.
For a good introduction to HHMs and parameter estimation see \citet{shumway1982, rabiner1986, thomas2014}.
In summary, each iteration of the EM algorithm starts with the E-step which calculates the forward and backward equations to get an estimate of the hidden variables given the entire sequence of observables.
The M-step then consists of maximising the expected log-posterior probability of the hidden parameters found in the E-step.
The M-step equations for LDS are given explicitly in \cite{shumway1982} therefore we do not provide them here.
We call these solutions the LDS-EM solutions.

\begin{algorithm}
\SetAlgoLined
\tcc{Initialise}
% Set $\Delta m_{\rm tol}$ to a convergence relative tolerance\;
Set $N_{\rm max}$ to a max number of iterations\;
Set $\bs{\Sigma}^{(0)} = \sigma^2 \bs{I}$ as an initial estimate of observational covariance of $\g$ per frequency\;
Set $\omega^{(0)}$ to an initial estimate of the DDTEC L\'evy variance per time-interval\;
% Set $m_0^{(0)}$ and $\gamma_0^{(0)}$ to initial prior mean and variance of DDTEC\;
Set $n=0$\;
 \While{$n < N_{\rm max}$}{
 \tcc{E-step}
 Set $i=0$\;
 \While{$i<T$
 \tcp*{Forward pass}
 }{
  Use $\omega^{(n)}$ and $\bs{\Sigma}^{(n)}$ to define $\mathcal{L}_i[q\mid\mid p]$\;
  Maximise $\mathcal{L}_i[q\mid\mid p]$ for $m_i^{(n)}$ and $\gamma_i^{(n)}$\;
  Set $i=i+1$\;
  }
  Set $i=T$\;
 \While{$i > 0$
 \tcp*{Backward pass}
 }{
  Revise estimate of $m_i^{(n)}$ and $\gamma_i^{(n)}$ with Rauch-recurrence relation\;
  Set $i=i-1$\;
  }
  \tcc{M-step}
  Set ${\omega^{(n+1)}}$ and  $\bs{\Sigma}^{(n+1)}$ to the LDS-EM solutions\;
%   \If{$\left|\frac{m^{(n+1)}_i - m^{(n)}_i}{m^{(n)}_i}\right| < \Delta m_{\rm tol} $ for all $i$}{
%   break\;
%   }
   Set $n=n+1$\;
 }
 \tcc{Final variational mean and variance}
 \Return{$\{m_i^{(n)}\}$, $\{\gamma_i^{(n)}\}$}
 \caption{Solving the recursive Bayesian estimation of DDTEC from Jones scalars problem using variational inference to approximate the analytically intractable forward distribution.
 It assumes the DDTEC is a L\'evy process with Gaussian steps.
 It uses the EM-algorithm to estimate the the observational covariance and L\'evy process step variance.}
 \label{alg:ddtec_inference}
\end{algorithm}

\begin{figure}
    \centering
    \includegraphics[width=\columnwidth]{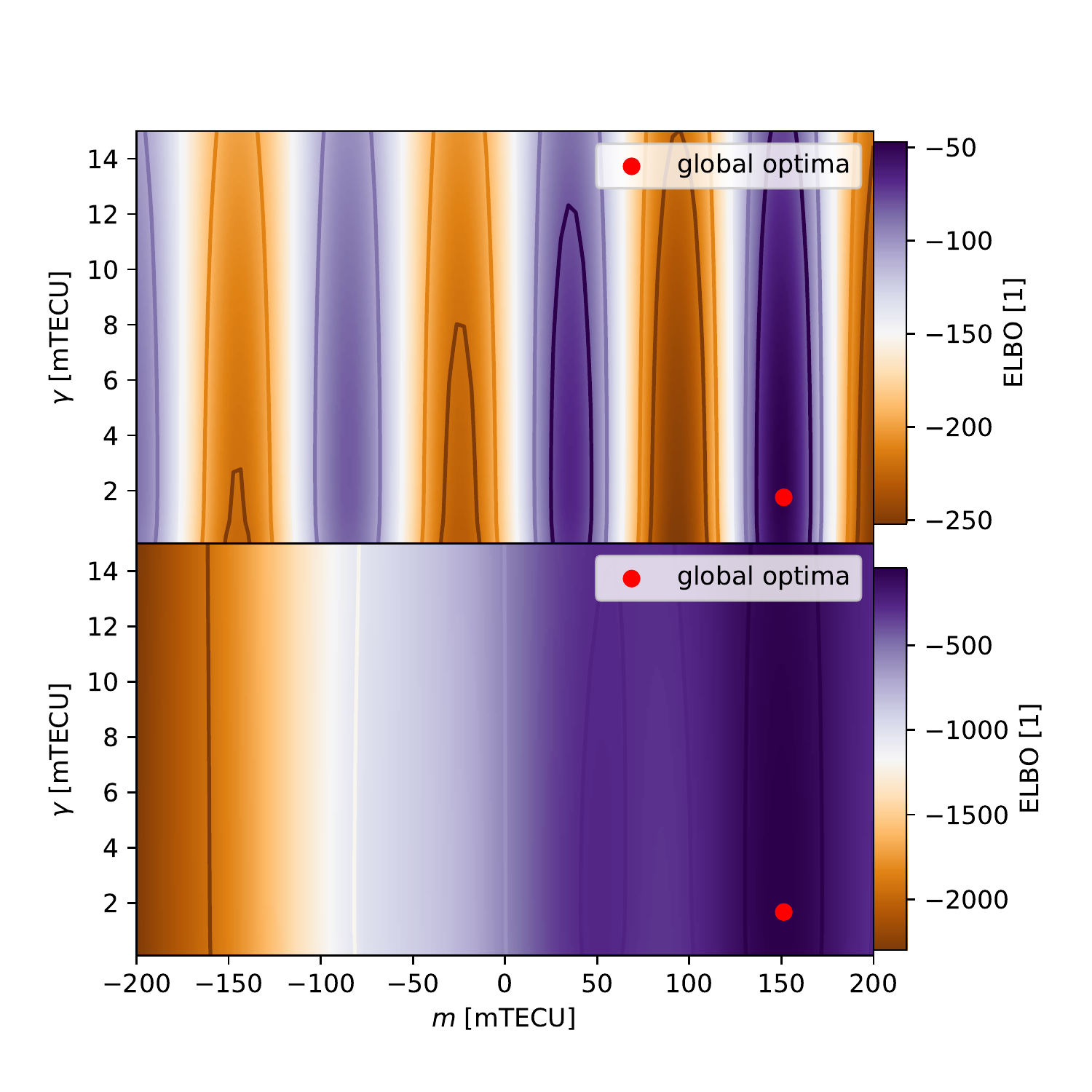}
    \caption{ELBO landscape of the simulated Jones scalars in \textbf{Figure}~\ref{fig:example_gains} during the first and second iterations of Algorithm~\ref{alg:ddtec_inference}. The top panel shows the ELBO basin during the first iteration and the lower panel shows the ELBO basin during the second iteration after one round of parameter estimation. The ground truth DDTEC is 150~mTECU in this example.}
    \label{fig:elbo_basin}
\end{figure}

\begin{figure}
    \centering
    \includegraphics[width=\columnwidth]{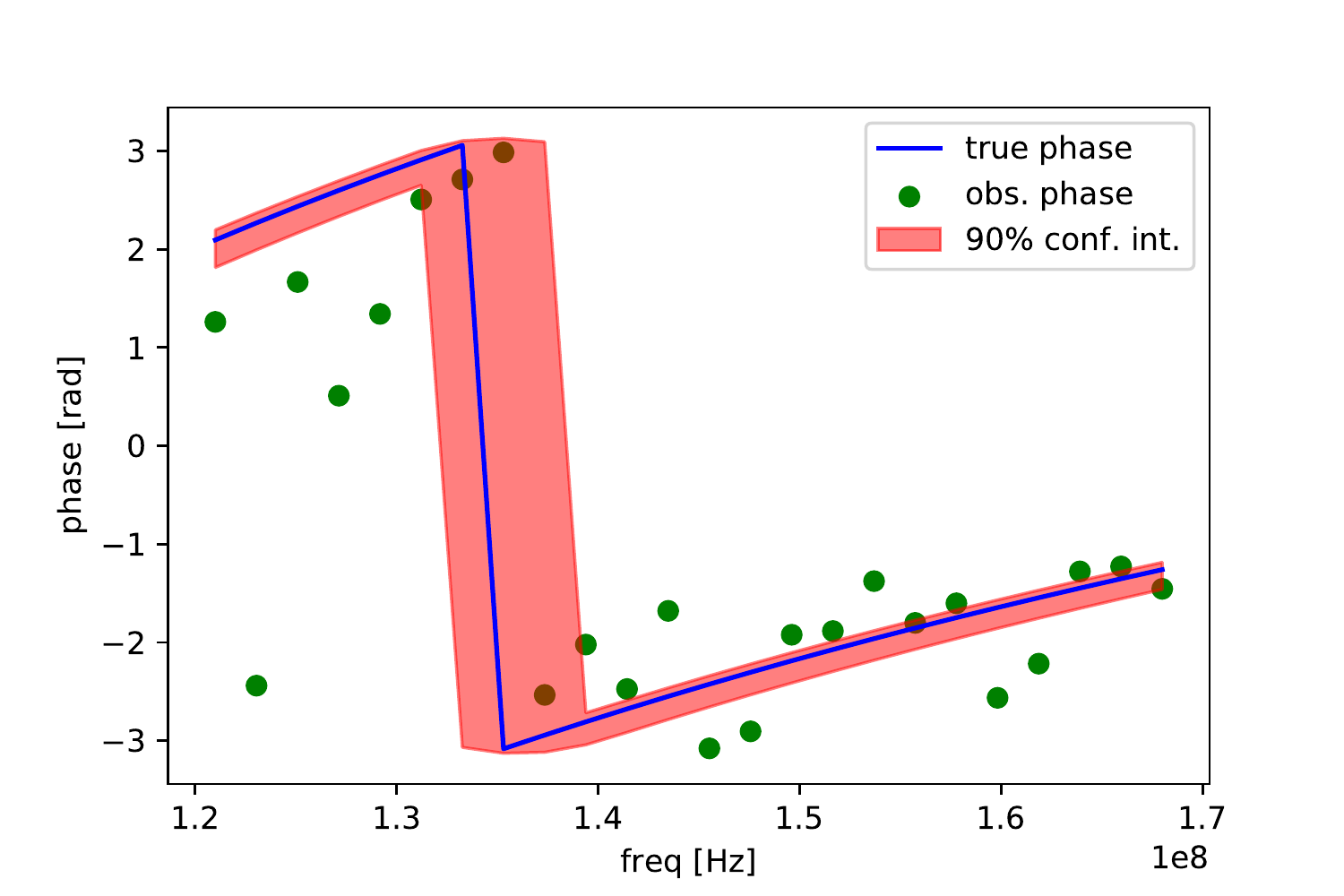}
    \caption{Posterior doubly differential phase solved with variational inference. The shaded region shows the central 90\% confidence interval following two passes of Algorithm~\ref{alg:ddtec_inference}. The resulting variational estimate for DDTEC is $149.1\pm 2.3$~mTECU. The ground truth DDTEC is 150~mTECU.}
    \label{fig:example_posterior_ddtec}
\end{figure}
The complete DDTEC inference algorithm is given in Algorithm~\ref{alg:ddtec_inference}.
For the initial prior DDTEC prior mean and prior uncertainty we use 0~mTECU and 300~mTECU respectively.
The top panel in \textbf{Figure}~\ref{fig:elbo_basin} shows the ELBO basin for the Jones scalar data in \textbf{Figure}~\ref{fig:example_gains} during the first forward pass iteration of the algorithm.
During the first iteration, the HMM parameters are not known and the ELBO basin is has many local optima. 
However, after the first iteration the HMM parameters are estimated and during the second pass most of the local optima are gone, corresponding to a better priors which were learned from the data. 
This happens because the constrained L\'evy process variance regularises the inferred DDTEC.
In situations with noisy data, but where \textit{a priori} the DDTEC changes very little, this can significantly improve the estimate of posterior DDTEC.
In situations where the DDTEC varies a lot between time steps, the time coupling regularisation helps less however it still reduces the variance of the estimates.
\textbf{Figure}~\ref{fig:example_posterior_ddtec} shows the resulting 90\% confidence interval of the posterior phase for this example simulated DDTEC inference problem, and it can be seen that the method correctly found the right phase wrap.

To globally optimise ELBO in the presence of many local optima, we use brute force to locate a good starting point, and then use BFGS quasi-Newton optimisation from that point to find the global optimum.
Because there are only two variational parameters the brute force grid search is very quick.

\subsubsection{Outlier flagging}

In this randomly chosen data set 0.1\% of the measured DDTEC are found to be outlier solutions -- approximately $1.6\times 10^3$ of $2.6\times 10^6$ optical pathways.
Outliers are visually characterised by a posterior uncertainty that is too small to explain its deviation from its neighbouring -- in time and direction -- DDTEC. 
Outliers drastically impact the performance of the screen inference step, since the uncertainties will be `trusted'.

We detect outliers using a two-part heuristic approach, where we value having more false-positives over false-negatives.
First, we filter for large jumps in DDTEC.
We use rolling-statistics sigma-clipping, where we flag DDTEC that deviate by more than 2 standard-deviations from the median in a rolling temporal window of 15~min.
Secondly, we filter for large jumps directionally.
We fit a smoothed multiquadric radial basis function to the DDTEC over direction and flag outliers where the DDTEC deviates by more than 8~mTECU.
Measured DDTEC which are flagged as outliers are given infinite uncertainty.

\subsection{Infer DDTEC screen}
\label{sec:infer_screen}

\begin{figure}
    \centering
    \includegraphics[width=\columnwidth]{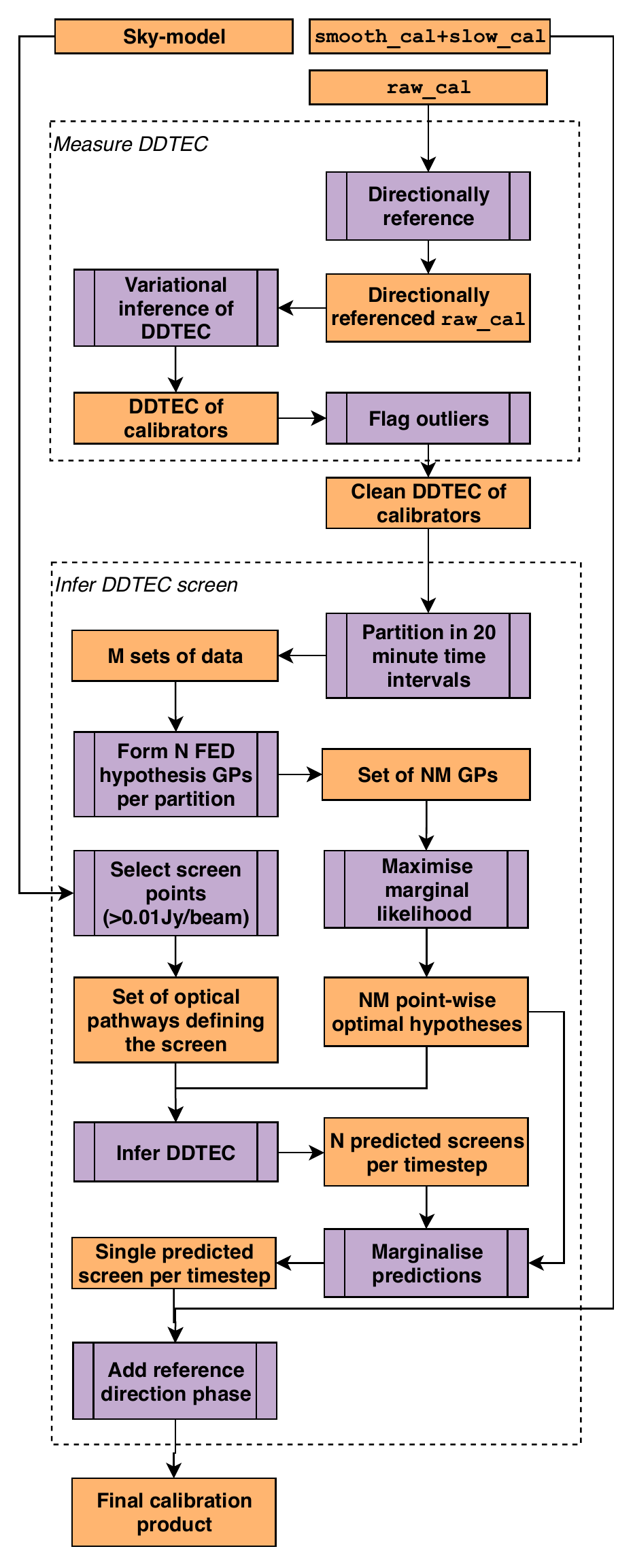}
    \caption{Flow diagram of the \textit{measure DDTEC} and \textit{infer DDTEC screen} steps. These steps can be considered the main difference between our method and the one used in the LoTSS-DR1 and DR2 pipelines.}
    \label{fig:measure_ddtec_and_infer}
\end{figure}

We will denote the DDTEC mean and uncertainties inferred with variational inference in Section~\ref{sec:vi_ddtec} as the measured DDTEC with associated measurement uncertainties.
In this section we show how we perform probabilistic inference of a DDTEC screen.
In order to differentiate the DDTEC screen from the measured DDTEC, we shall call them the inferred DDTEC.
The process is visualised in the correspondingly labelled box in \textbf{Figure}~\ref{fig:measure_ddtec_and_infer}.

We define the screen directions by selecting the brightest 250 directions with apparent brightness greater than 0.01~$\mathrm{Jy}\,\mathrm{beam}^{-1}$ and separated by at least {4\arcmin}.
We exclude the original calibrator directions from the set of screen directions, which already have optimal solutions provided in \texttt{smooth\_cal+slow\_cal}.

We model the measured DDTEC over optical pathways, indexed by $(\x, \k)$, following the ionospheric model proposed in \citet{2020A&A...633A..77A} which geometrically models the ionosphere as a flat, thick, infinite-in-extent layer with free electron density (FED) realised from a Gaussian process.
Whereas in \citet{2020A&A...633A..77A} the modelled quantity was differential total electron content (DTEC), here we model DDTEC which requires a modification to their model.
Generalising the procedure outlined therein, we can arrive at similar expressions as their Eqs.~11 and 14.
Since the results can be arrived at by following the same procedure, we do not show the derivation here and simply state the DDTEC prior mean and covariance,
\begin{align}
     m_{\ddtec}(\x, \k) =& \mathbb{E}[\Delta_0^2\tau(\x, \k)] = 0,\label{eq:mean_a}\\
    K_{\ddtec}((\x_i, \k), (\x_j, \k')) = & \mathbb{E}[\Delta_0^2\tau(\x_i, \k) \Delta_0^2\tau(\x_j, \k')]\\
    =& 
    I_{ij}^{{\vec{k}}{\vec{k}'}} 
    -I_{i0}^{{\vec{k}}{\vec{k}'}}
    -I_{ij}^{{\vec{k}}{\vec{k}}_0}
    +I_{i0}^{{\vec{k}}{\vec{k}}_0}\notag\\
    -&I_{0j}^{{\vec{k}}{\vec{k}'}} 
    +I_{00}^{{\vec{k}}{\vec{k}'}}
    +I_{0j}^{{\vec{k}}{\vec{k}}_0}
    -I_{00}^{{\vec{k}}{\vec{k}}_0}\notag\\
   -&I_{ij}^{{\vec{k}}_0{\vec{k}'}} 
    +I_{i0}^{{\vec{k}}_0{\vec{k}'}}
    +I_{ij}^{{\vec{k}}_0{\vec{k}}_0}
    -I_{i0}^{{\vec{k}}_0{\vec{k}}_0}\notag\\
    +&I_{0j}^{{\vec{k}}_0{\vec{k}'}}
    -I_{00}^{{\vec{k}}_0{\vec{k}'}}
    -I_{0j}^{{\vec{k}}_0{\vec{k}}_0}
    +I_{00}^{{\vec{k}}_0{\vec{k}}_0},\label{eq:cov_a}
\end{align}
where we have used the same notation with latin subscripts denoting antenna indices.
Each term in Eq.~\ref{eq:cov_a} is a double integral
\begin{align}
    I_{ij}^{{\vec{k}}{\vec{k}}'} 
    =& \int_{s_i^{{\vec{k}}-}}^{s_i^{{\vec{k}}+}} \int_{s_j^{{\vec{k}}'-}}^{s_j^{{\vec{k}}'+}} K\left( \x_i + s{\k},\x_j + s'{\k}'\right)\,\mathrm{d}s\mathrm{d}s',\label{eq:I_a}
\end{align}
of the FED covariance function $K$, and the integration limits are given by,
\begin{align}
s_i^{{\vec{k}}\pm} = \left(a \pm \frac{b}{2} - \left(\x_i - \x_0\right) \cdot \hat{\vec{z}}\right) (\k\cdot\hat{\vec{z}})^{-1},
\end{align}
where $a$ is the height of the centre of the ionosphere layer above the reference antenna, and $b$ is the thickness of the ionosphere layer.
The coordinate frame is one where the $\hat{\vec{z}}$ is normal to the ionosphere layer, e.g. the East-North-Up frame.

The FED kernel, $K$, that is chosen determines the characteristics of the resulting DDTEC.
In Section~\ref{sec:adaptive} we discuss our choice of FED kernels.
For a comprehensive explanation of the effect of different FED kernels on DTEC, and examples of inference with this model see \citet{2020A&A...633A..77A}.
Importantly, the only difference between inference of DDTEC and inference of DTEC is in the number of terms in Eq.~\ref{eq:cov_a}, where there are 16 for DDTEC and only 4 for DTEC. 

Using Eqs.~\ref{eq:mean_a} and \ref{eq:cov_a} to define the DDTEC prior distribution, the posterior distribution for the DDTEC screen given our measured DDTEC is inferred following standard Gaussian process regression formulae \citep[e.g.][]{Rasmussen:2005:GPM:1162254}.
Specifically, given a particular choice for the FED kernel, $K$, if the set of optical pathways of the measured DDTEC are given by $\Z$, and the set of all optical pathways that we wish to infer DDTEC over are given by $\Z^*$ then the posterior distribution is \citep[\textit{cf.} Eq.~19 in][]{2020A&A...633A..77A},
\begin{align}
    P\left(\ddtec(\Z^*) \mid \ddtec(\Z), K\right)
    =&\mathcal{N}[\ddtec(\Z^*) \mid \vec{m}(\Z^*), \vec{K}(\Z^*, \Z^*)]
\end{align}
where the conditional mean is $\vec{m}(\Z^*) = K_{\ddtec}(\Z^*,\Z)\bs{B}^{-1}\ddtec(\Z)$, the conditional covariance is $\vec{K}(\Z^*, \Z^*) = K_{\ddtec}(\Z^*,\Z^*) - K_{\ddtec}(\Z^*,\Z)\bs{B}^{-1}K_{\ddtec}(\Z,\Z^*)$, and $\bs{B} = K_{\ddtec}(\Z,\Z) + \Sigma_{\ddtec}(\Z,\Z)$. 
The measured DDTEC mean and uncertainties are $\ddtec(\Z)$ and $\Sigma_{\ddtec}(\Z,\Z)$ respectively.
To handle infinite measurement uncertainties, one factors $\Sigma_{\ddtec}(\Z,\Z)$ into the product of two diagonal matrices, then symmetrically factors the diagonals out of $\bs{B}$.
The inverted infinities then act to zero-out corresponding columns on right multiplication and rows on left multiplication.

For a given choice of FED kernel, $K$, the geometric parameters of the ionosphere and FED kernel hyper parameters are determined by maximising the log-marginal likelihood of the measured DDTEC \citep[\textit{cf.} Eq.~18 in][]{2020A&A...633A..77A},
\begin{align}
     \log P\left(\ddtec(\Z) \mid K\right)
    =& \log \mathcal{N}[\ddtec(\Z) \mid 0,  \bs{B}].\label{eq:lml_a}
\end{align}
% When computing the log-marginal likelihood of the measured DDTEC, the log of the determinant of $\bs{B}$ will have zero-times-infinity terms corresponding to flagged outliers.
% Applying l'H\^opital's rule these terms become zero.

For all our Gaussian process computations we make use of GPFlow \citep{GPflow2017}, which uses auto-differentiation of Tensorflow \citep{tensorflow2015} to expedite complex optimisation procedures on both CPUs and GPUs.

\subsubsection{Accounting for a dynamic ionosphere}
\label{sec:adaptive}
The ionosphere is very dynamic, with a variety of influencing factors.
It can change its over-all behaviour in a matter of minutes \citep[e.g.][]{mevius2016, 2017MNRAS.471.3974J}.
In our experimentation, we found that choosing the wrong FED kernel results in very poor quality screens due to a systematic modelling bias.

We delimit two cases of wrongly specifying the FED kernel.
The first is when the true FED covariance is stationary but the spectral properties are wrongly specified.
This can happen, for example, when the character of the ionosphere changes from rough to smooth, or vice-versa.
The tomographic nature of $K_{\ddtec}$ then falsely infers DDTEC structure that is not there.
In the second case, the true FED covariance is non-stationary and a stationary FEED kernel is assumed.
This happens, for example, when a travelling ionospheric disturbance (TID) passes over the field of view \citep{tol2009}.
In this case, the FED has a locally non-stationary component and the tomographic nature of the DDTEC kernel will falsely condition on the TID structures causing erratic predictive distributions.

Both types of bias, can be handled by a dynamic marginalisation over model hypotheses.
To do this, we denote each choice of FED kernel as a hypothesis.
For each choice of FED kernel, we optimise the FED and ionosphere hyper parameters by maximising the log-marginal likelihood of the measured DDTEC, Eq.~\ref{eq:lml_a}.
We then define the hypothesis-marginalised posterior as,
\begin{align}
    P\left(\ddtec(\Z^*) \mid \ddtec(\Z)\right) = \sum_i w_i P\left(\ddtec(\Z^*) \mid \ddtec(\Z), K_i\right),
\end{align}
where $w_i = P\left(\ddtec(\Z) \mid K_i\right) / \sum_i P\left(\ddtec(\Z) \mid K_i\right)$.
This marginalisation procedure can be viewed as a three-level hierarchical Bayesian model, \citep[e.g. chapter 5 in][]{Rasmussen:2005:GPM:1162254}.
The hypothesis marginalised distribution is a mixture of Gaussian distributions, which is a compound model and is no longer Gaussian in general.
We approximate the mixture with a single Gaussian, by analytically minimising the Kullbeck-Leibler divergence from the mixture to the single Gaussian. 
For reference, the mean of the single Gaussian approximation is the convex sum of means weighted by the hypothesis weights. 

For our FED hypothesis kernels we select the Mat\'ern-$p/2$ class of kernels, for $p\in\{1,3,5, \infty\}$, and the rational-quadratic kernel \citep{Rasmussen:2005:GPM:1162254}.
All selected kernels are stationary.
While we could incorporate non-stationary FED kernel hypotheses, to handle the second type of bias, we do not explore such kernels in this work.
See \citet{2020A&A...633A..77A} for examples of the Mat\'ern-$p/2$ class of FED kernels.

\subsubsection{Computational considerations}
\label{sec:comp_considerations}
In order for this approach to be practically feasible, the computation should only require a few hours. 
The main bottleneck of the computation is in computing the 16 double integral terms in Eq.~\ref{eq:cov_a} for every pair of optical pathways, and inverting that matrix.
A single covariance matrix requires evaluating $16(N_{\rm res} N_{\rm ant} N_{\rm cal})^2$ double precision floating point numbers which are stored in a $ N_{\rm ant} N_{\rm cal} \times  N_{\rm ant} N_{\rm cal}$ matrix.
In the case of P126+65, using a abscissa resolution of $N_{\rm res}=5$ as suggested by \citep{2020A&A...633A..77A}, the covariance matrix is approximately 23GB of memory, with approximately 600GB of memory used in intermediate products.
The large intermediate products means that the covariance matrix evaluation is memory bus speed bound.
Furthermore, the inversion scales with $N_{\rm ant}^3 N_{\rm cal}^3$.
Hyper parameter optimisation requires approximately 200 iterations of hessian-based gradient descent, with each iteration computing the matrix and inverse at least once.
Therefore, a lower limit on the number of matrix inversions required for a typical observation with 5 FED kernel hypotheses is $10^6$.
On a single computer with 32 cores and 512 GB of RAM this takes several weeks.
Chunking data in time produces modest savings.
Using GPUs would make this calculation feasible in just a few hours.
However, while GPUs are intrinsically enabled by Tensorflow, we do not currently have access to GPUs with enough memory.

To do this scalably with enough FED kernel hypotheses to handle the dynamic nature of the ionosphere, would require too much computing power for a single measurement set calibration, and, unfortunately, a trade-off must be made.
Therefore, we first make the approximation that the ionosphere is thin, i.e. we let $b \to 0$.
When we make this approximation the double integral terms become,
\begin{align}
    \tilde{I}_{ij}^{{\vec{k}}{\vec{k}}'} 
    =& K\left( \x_i + s_i^{\k0}{\k},\x_j + s_j^{\k'0}{\k}'\right)\label{eq:I_b}
\end{align}
where $s_i^{\k0} = \left(a  - \left(\x_i - \x_0\right) \cdot \hat{\vec{z}}\right) (\k\cdot\hat{\vec{z}})^{-1}$.
This reduces the memory size of intermediate products by $N_{\rm res}^2$, approximately from 600GB to 30GB, however the size of the matrix is still still the same, and the same problem of inverting a large matrix many times appears.

Therefore, we introduce a second approximation, that there is no \textit{a priori} coupling between antennas.
When we do this, Eq.~\ref{eq:cov_a} becomes, 
\begin{align}
    K^{\x_i}_{\ddtec}(\k, \k') =&
    \tilde{I}_{ii}^{{\vec{k}}{\vec{k}'}} 
    -\tilde{I}_{i0}^{{\vec{k}}{\vec{k}'}}
    -\tilde{I}_{ii}^{{\vec{k}}{\vec{k}}_0}
    +\tilde{I}_{i0}^{{\vec{k}}{\vec{k}}_0}\notag\\
    -&\tilde{I}_{0i}^{{\vec{k}}{\vec{k}'}} 
    +\tilde{I}_{00}^{{\vec{k}}{\vec{k}'}}
    +\tilde{I}_{0i}^{{\vec{k}}{\vec{k}}_0}
    -\tilde{I}_{00}^{{\vec{k}}{\vec{k}}_0}\notag\\
   -&\tilde{I}_{ii}^{{\vec{k}}_0{\vec{k}'}} 
    +\tilde{I}_{i0}^{{\vec{k}}_0{\vec{k}'}}
    +\tilde{I}_{ii}^{{\vec{k}}_0{\vec{k}}_0}
    -\tilde{I}_{i0}^{{\vec{k}}_0{\vec{k}}_0}\notag\\
    +&\tilde{I}_{0i}^{{\vec{k}}_0{\vec{k}'}}
    -\tilde{I}_{00}^{{\vec{k}}_0{\vec{k}'}}
    -\tilde{I}_{0i}^{{\vec{k}}_0{\vec{k}}_0}
    +\tilde{I}_{00}^{{\vec{k}}_0{\vec{k}}_0}\label{eq:cov_b}
\end{align}

This covariance matrix requires only computing $16(N_{\rm cal})^2$ double precision floating point numbers and requires inverting an $N_{\rm cal} \times N_{\rm cal}$ matrix.
This is extremely fast, even with CPUs and the whole P126+65 measurement set can be done in 50 minutes with 5 FED kernel hypotheses.

After making these approximations with a stationary FED kernel it becomes difficult to infer the height of the ionosphere, $a$.
Rather, if the FED kernel is isotropic and parametrised by a length scale, say $l$, then the marginal likelihood becomes sensitive to the ratio $a/l$.
Therefore, it is still possible to learn something of the ionosphere with this approximation.

\subsection{Image screen}

DDFacet has internal capability for applying solutions during imaging in arbitrary directions \citep{2018A&A...611A..87T}.
DDFacet typically works with its own proprietary solution storage format.
To facilitate working with our solutions we extended DDFacet to work with, and get the directions layout from, H5Parm files whose specifications are defined by LoSoTo \citep{2019A&A...622A...5D}.

We convert the \texttt{smooth\_cal+slow\_cal} solutions to H5Parm format and then convert DDTEC to DTEC by adding back the reference direction phase from \texttt{smooth\_cal}. 

We use DDFacet's hybrid matching pursuit clean algorithm, with 5 major cycles, $10^6$ minor cycles, a Briggs robust weight of -0.5 \citep{briggs}, a peak gain factor of 0.001, and clean threshold of $100~\mu\mathrm{Jy}\,\mathrm{beam}^{-1}$.
We initialise the clean mask with the final mask from LoTSS-DR2.
All other settings are the same as in the LoTSS-DR2 pipeline and can be found in \citet{2019A&A...622A...1S}.

We also apply the same settings to image the solutions in \texttt{smooth\_cal+slow\_cal} for comparison purposes later.

\section{Results}
\label{sec:results}

\begin{figure*}
    \centering
    \includegraphics[width=\textwidth]{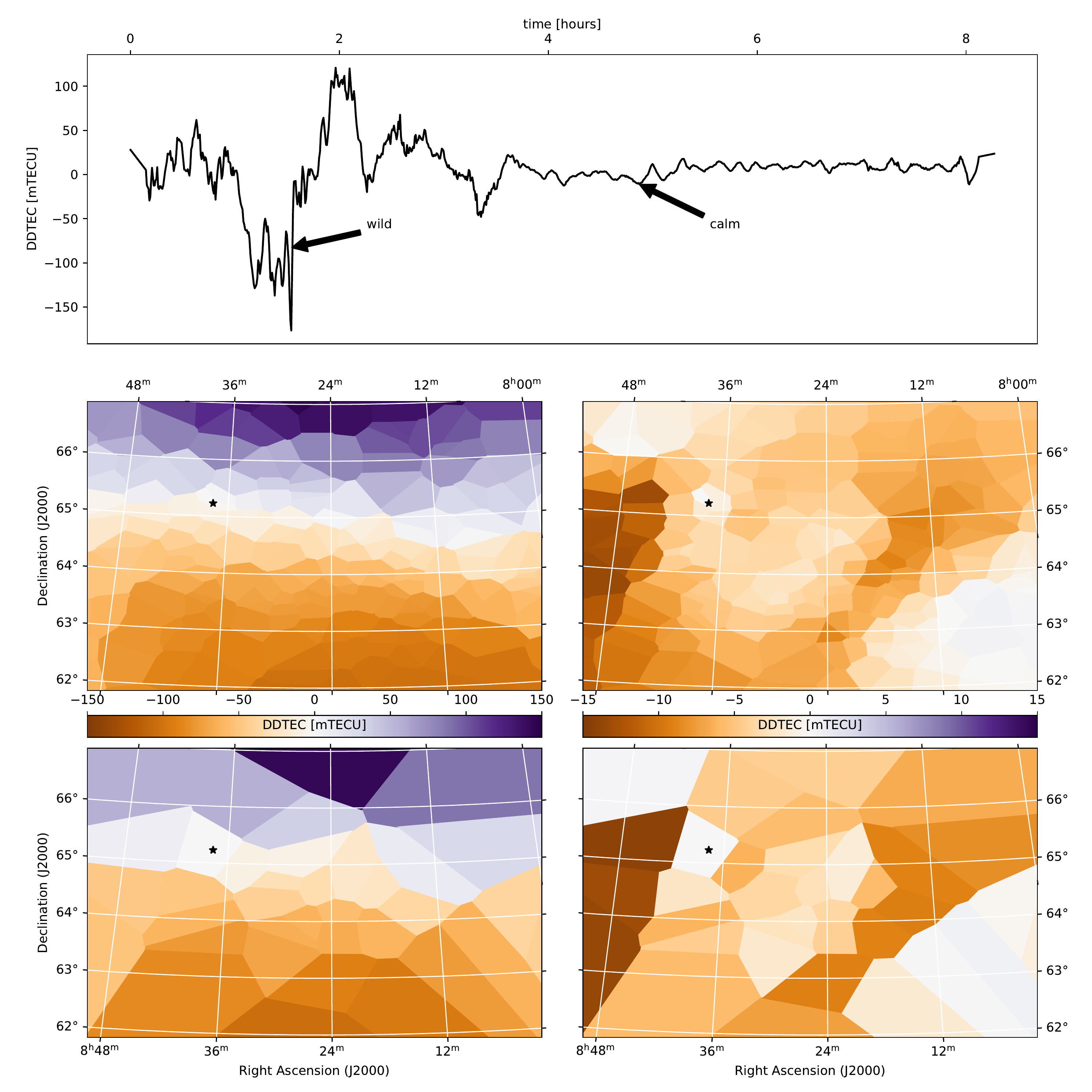}
    \caption{DDTEC-screens and temporal profile of LOFAR antenna RS508HBA.
    The lower four plots show what the path-length distortions due to the ionosphere looked like from the perspective of antenna RS508HBA during wild and calm periods.
    The middle row shows the inferred DDTEC-screen, and the bottom row shows the measured DDTEC-screen.
    The black star indicates the reference direction, hence the DDTEC is always zero in that direction.
    The top plot shows the DDTEC for a single direction over the course of the observation, and clearly shows that the temporal behaviour changed around 3.5 hours into the observation.}
    \label{fig:ddtec_panel}
\end{figure*}

\textbf{Figure}~\ref{fig:ddtec_panel} shows several dimensional slices of the resulting DDTEC screens for LOFAR remote antenna RS508HBA, which is 37~km from the reference antenna.
Remote antennas are typically more difficult to calibrate due to the drop-off of flux on smaller angular scales (longer baselines), therefore they are a good choice for showing the performance of the DDTEC measurement and screen inference.
The top panel shows the temporal evolution of a single direction.
We observe two distinct ionospheric behaviours over the course of the observation.
From 0 to 3.5 hours the ionosphere is `wild' with DDTEC variations greater than 150~mTECU, and the variation between time steps more noise-like.
From 3.5 hours until the end of the observation the DDTEC rarely exceeds 10~mTECU, and the variations between time steps are smoother.
For comparision, a central antenna within 500~m of the reference antenna typically has DDTEC variations of 10~mTECU.
In order to have such a small DDTEC so far from the reference antenna, the FED of the ionosphere should be highly spatially correlated on length scales longer than that baseline.
This implies that a structure larger than 50~km, potentially at low altitude, is passing over the array during that time.

Since we applied FED model hypothesis marginalisation, we can confirm that the spatial power spectrum indeed changed during these two intervals.
During the first half of the observation, coinciding with the `wild' ionosphere, the most highly weighted FED kernel hypothesis was the Mat\'ern-$1/2$ kernel, followed by the Mat\'ern-$3/2$ kernel.
During the last half of the observation, coinciding with the `calm' ionosphere, the most highly weighted FED kernel hypothesis was the Mat\'ern-$3/2$ kernel, followed by the Mat\'ern-$5/2$ kernel.

The change in temporal variation roughness implies that the temporal power spectrum of the FED changed in shape, becoming more centrally concentrated in the last half of the observation.
If we apply the frozen flow assumption, then, over short enough time intervals, the temporal covariance becomes the FED spatial covariance and therefore we expect to see rougher temporal correlations when the spatial correlations become rougher.
The frozen flow assumption is thus supported by the fact that the first and last half of the observations are better described by rough and smooth FED kernels, respectively, matching the temporal behaviour.

The middle and bottom rows of \textbf{Figure}~\ref{fig:ddtec_panel} show the antenna-based DDTEC-screens -- showing what the ionosphere looks like from the perspective the antenna -- of our inferred DDTEC and measured DDTEC respectively.
The left columns are a slice during the `wild' interval and the right column is a slice during the `calm' interval (points indicated in the temporal profile). 
Note that during the `calm' time the DDTEC-screens are mostly negative.
From Equation~\ref{eq:mean_a} we expect that the DDTEC should be zero on average, which seems at odds with the overall negative DDTEC-screen at this time slice.
As pointed out in \citet{2020A&A...633A..77A}, the mean DTEC, and therefore mean DDTEC, should be zero only in the case of a flat infinite in extend ionosphere layer, and than in a model where the ionosphere follows the curvature of the Earth the mean is not zero.
Indeed, the remote antennas of LOFAR are far enough from the reference antenna to have non-zero mean DDTEC due to curvature.

Since we have applied the two approximations in Section~\ref{sec:comp_considerations}, our model is no longer tomographic therefore we should not expect super-resolution DDTEC inference as observed in \citet{2020A&A...633A..77A}.
However, our model still retains the directional coupling of a thin-layer ionosphere model.
Therefore, we observe that the reference direction is constrained to have zero DDTEC in the inferred DDTEC screens.

\begin{figure*}
    \centering
    \includegraphics[width=\textwidth]{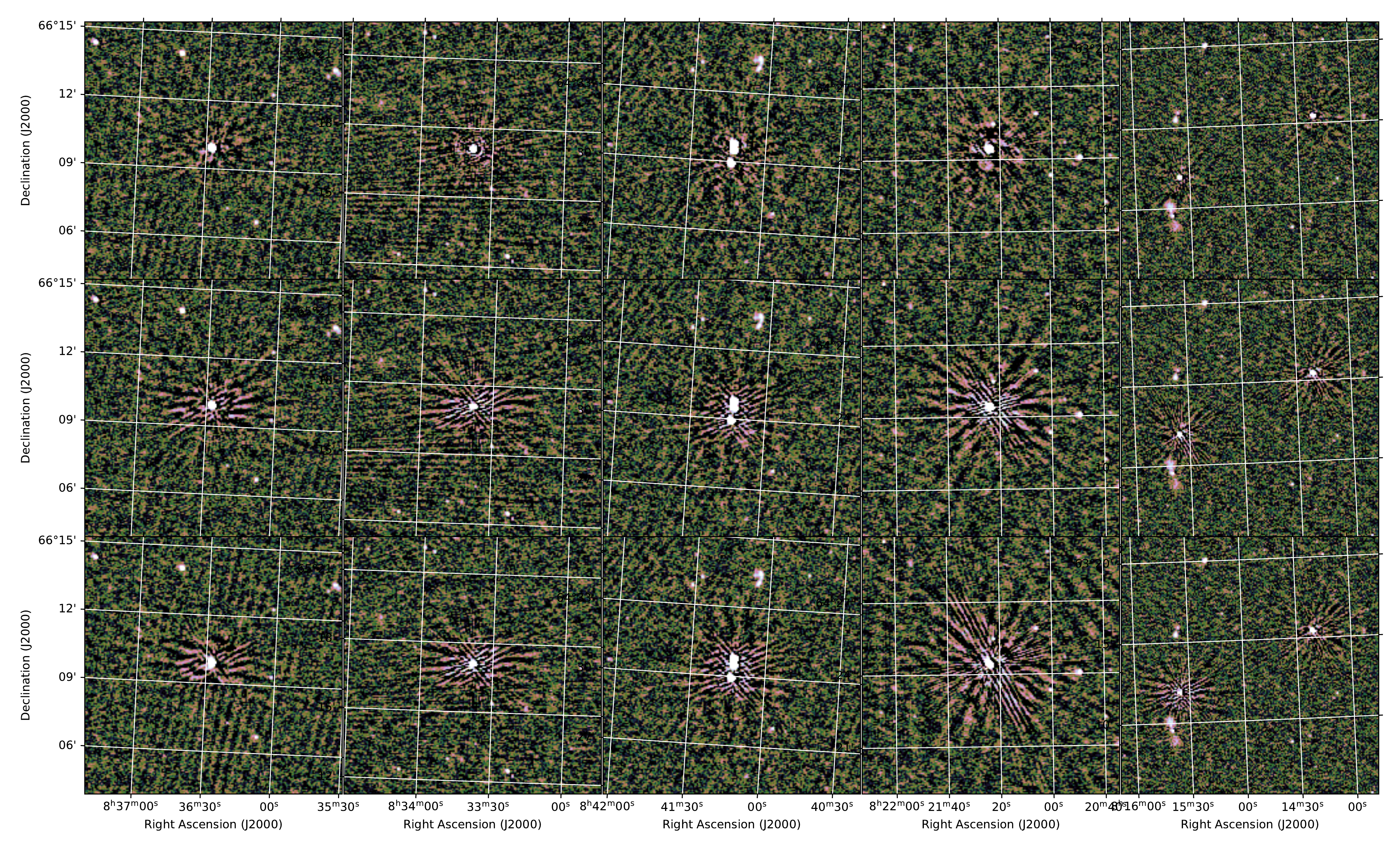}
    \caption{Comparison of direction dependent effects in regions far from a calibrator.
    The bottom row shows the archival LoTSS-DR2 image, the middle row shows the image when applying \texttt{smooth\_cal+slow\_cal}, and the top row shows the final image with our inferred DDTEC screen applied.
    The column numbers correspond to the regions labelled in \textbf{Figure}~\ref{fig:field_regions}.}
    \label{fig:image_panel}
\end{figure*}

\textbf{Figure}~\ref{fig:image_panel} shows, side-by-side, several sources in the LoTSS-DR2 archival image, the \texttt{smooth\_cal+slow\_cal} image, and the inferred DDTEC screen image.
The column numbers correspond to the region indicated in \textbf{Figure}~\ref{fig:field_regions} with red circles, which shows that all selected sources are between calibrators.
The LoTSS-DR2 and \texttt{smooth\_cal+slow\_cal} sources display similar dispersive phase errors, which implies that the isoplanatic assumption was violated in that region of the image as some point during the observation.
In the inferred DDTEC screen image the dispersive phase error effects appear less pronounced, indicating that the inferred DDTEC screen provided a better calibration than nearest-neighbour interpolation of the calibrator solutions. 

\begin{figure}
    \centering
    \includegraphics[width=\columnwidth]{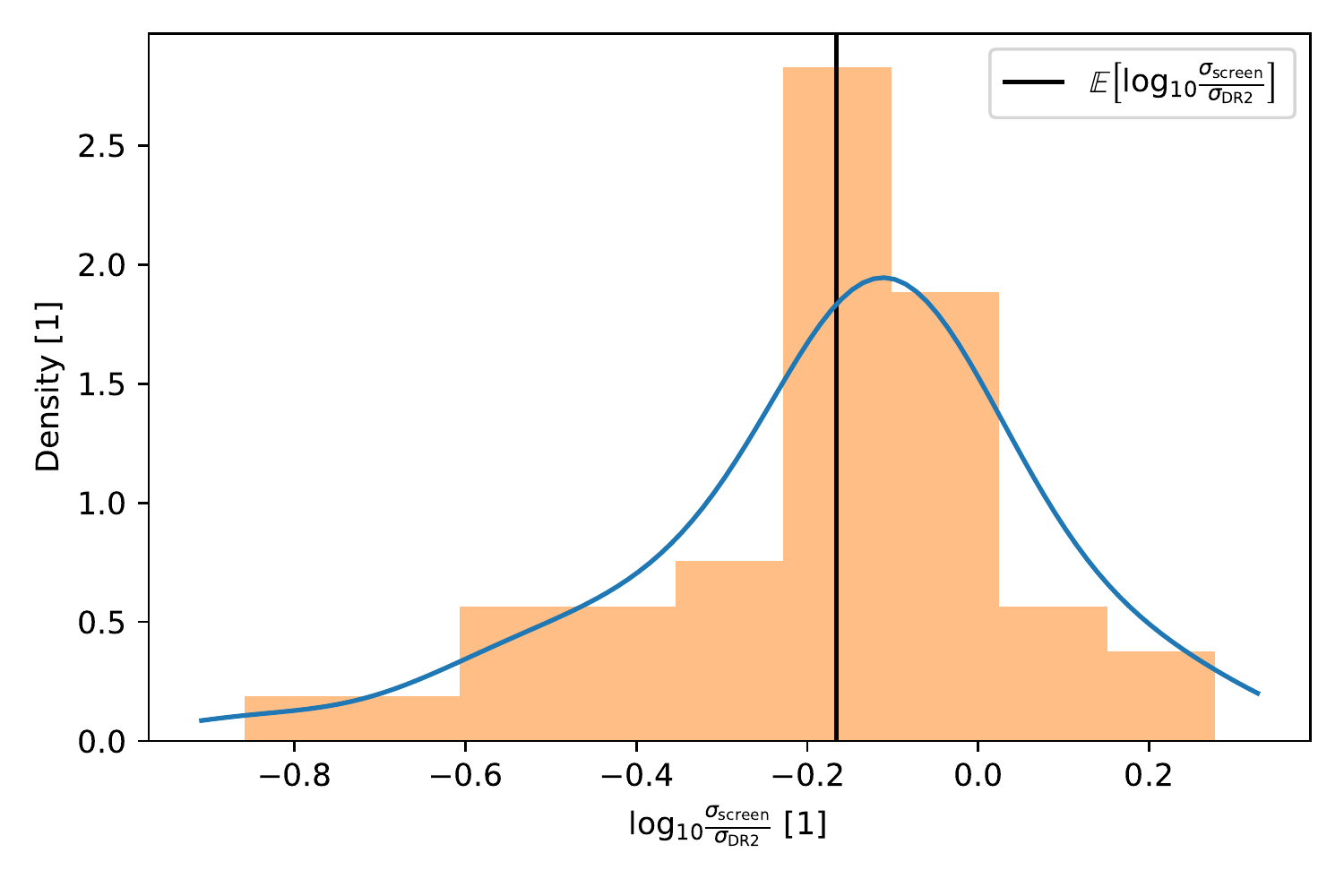}
    \caption{Histogram and Gaussian kernel density estimation of the difference in magnitude of root-mean-square residuals in annuli around bright source -- screen corrected, $\sigma_{\rm screen}$, over LoTSS-DR2 archival, $\sigma_{\rm DR2}$.
    The root-mean-square residuals are calculated within an annulus between 12\arcsec to 90\arcsec centred on sources brighter than $0.1~\mathrm{Jy}\,\mathrm{beam}^{-1}$.
    The residuals are primary-beam uncorrected.
    The black line denotes the mean ratio, and corresponds to $\sigma_{\rm screen}=0.68\sigma_{\rm DR2}$.}
    \label{fig:dynamic_range}
\end{figure}

In order to quantify the improvement to the image quality caused by applying the inferred DDTEC screen we systematically compare the scale of artefacts in the vicinity around bright sources within the primary beam.
We consider an annulus from 12\arcsec to 90\arcsec around all sources brighter than 0.1~$\mathrm{Jy}\,\mathrm{beam}^{-1}$, and calculate the root-mean-squared residuals for the LoTSS-DR2 image, $\sigma_{\rm DR2}$, and inferred DDTEC screen image, $\sigma_{\rm screen}$, within that annulus.
\textbf{Figure}~\ref{fig:dynamic_range} shows the histogram of $\log \sigma_{\rm screen} / \sigma_{\rm DR2}$.
In total, 76\% of the sources have $\sigma_{\rm screen} < \sigma_{\rm DR2}$ with a mean reduction in $\sigma_{\rm screen}$ of 32\% with respect to $\sigma_{\rm DR2}$.

% The artefacts around a source depend upon the brightness of the source, which therefore prompts a measure of dynamic range equal to the ratio of peak source brightness to the root-mean-squared residuals in the annulus.
% In the LoTSS-DR2 archival image the maximum so-defined dynamic range was $4.5\times 10^3$ and in the inferred DDTEC screen image the maximum dynamic range is $6.5\times 10^3$.

\section{Discussion}

In P126+65 there is a clear improvement over LoTSS-DR2 resulting from applying our inferred DDTEC screens.
Remarkably, we see this improvement using only 35 in-field calibrators compared to the 46 used in LoTSS-DR2.
This suggests that the method may have applications to other regimes where calibrators are very sparse, such as in very-long baseline interferometry or at ultra-low frequencies.

However, it is unclear if this observation is typical and if the same method applied to other observations would achieve similar results.
One obvious difference between observations is the different layout of sources, and therefore potential calibrators.
The distribution of bright sources in P126+65 is relatively uniform over the field of view which plays to the advantage of our screen inference method.
Due to the approximations in Section~\ref{sec:comp_considerations} the method is no longer tomographic and sparsity of calibrators could have a significant impact on screen quality.
Therefore, one of the next steps would be to test the performance on a large number of observations.
Furthermore, there are potential ways in which to make the tomographic approach computationally efficient and therefore computationally viable, however these require more research.

There is also the question of how robust this method is to various systematics.
For example, it is not yet known if unsubtracted compact, or diffuse, emission could be absorbed during the calibration, as is known to occur if there are too many degrees of freedom during calibration.
Another source of uncertainty is the robustness of this method to the behaviour of the ionosphere.
In this observation there were two distinct characters of ionosphere observed; one `wild' and one 'calm'.
There are perhaps many more types of ionospheres that could be encountered \citep[e.g.][]{mevius2016,2017MNRAS.471.3974J}.
Due to the FED hypothesis marginalisation, we expect that any new ionosphere should be manageable -- so long as the calibrators are not too sparse -- by incorporating the right FED kernel hypotheses.

Furthermore, while the changing ionosphere character may be handled with model weighting, the quality of the inferred DDTEC screens ultimately depends upon the quality of the measured DDTEC.
If this method were to be applied to LOFAR-LBA or long baseline data, the robustness of the DDTEC measurements must be guaranteed. 
For P126+65 0.1\% of all measured DDTEC were found to be outliers, however this could easily be much larger when there is less flux in the field against which to calibrate -- as happens at lower frequencies and on very long baselines -- or when the ionosphere's coherence time is shorter than the time interval of calibration (tens of seconds).

The improvements to root-mean-squared residuals near bright sources would be more appreciable with a lower thermal noise.
Therefore, an interesting avenue to test this method on would be on multi-epoch observations with tens to hundreds of hours of data.
In a deep observation, scattering of emission off of the ionospheric can cause low level structure in the image mimicing astrophysical sources, e.g. cosmic web accretion shocks hidden below the thermal noise.
Any wide-field statistical study of faint emission should ensure to properly calibrate such effects.

\section{Conclusion}

In this paper we have put forward a method for DD calibration and imaging of wide-field low-frequency interferometric data by probabilistically inferring DDTEC for all bright sources (>0.01$\mathrm{Jy}\,\mathrm{beeam}^{-1}$) in the field of view, using a physics-informed model.
In order to do this, we propose a HMM with variational inference of the forward distribution for measuring DDTEC with uncertainties from Jones scalars solved against isolated calibrators.
Isolating the calibrators was found to be important as this gives the Jones scalars well-defined effective directions.
We handle the dynamic nature of the ionosphere by marginalising the probabilistic model over a number of FED kernels hypotheses.
We only explore stationary hypothesis kernels, therefore this method may not perform well on ionospheres with TIDs.

We tested the method on a randomly selected observation taken from the LoTSS-DR2 archive.
The resulting image had fewer direction dependent effects, and the root-mean-squared residuals around bright sources were reduced by about 32\% with respect to the LoTSS-DR2 image.
Remarkably, we achieved this improvement using only 35 calibrator directions compared to 46 used in LoTSS-DR2.

While these results are promising, the robustness of the method must be verified on more observations.

\begin{acknowledgements}
J.\ G.\ A. and H.\ T.\ I. acknowledge funding by NWO under `Nationale Roadmap Grootschalige Onderzoeksfaciliteiten', as this research is part of the NL SKA roadmap project.
J.\ G.\ A. and H.\ J.\ A.\ R. acknowledge support from the ERC Advanced Investigator programme NewClusters 321271. R.\ J.\ vW. acknowledges support of the VIDI research programme with project number 639.042.729, which is financed by the Netherlands Organisation for Scientific Research (NWO). 
\end{acknowledgements}

\bibliographystyle{aa}
\bibliography{cite}

\onecolumn
\clearpage
\appendix

\section{Factoring commutative DI dependence from the RIME}
\label{app:ddtec}

Let us consider the effect of directionally referencing spatially referenced commutative Jones scalars. 
In particular, we'll look at phases of the Jones scalars, and assume amplitudes of one, however the same idea extends to amplitude by considering log-amplitudes which can be treated like a pure-imaginary phase.

Let $g(\x, \k)=e^{\iota \phi(\x, \k)}$ be a Jones scalar, and consider the necessarily non-unique decomposition of phase into DD and DI components,
\begin{align}
\phi(\x, \k) =  \phi^{\rm DD}(\x, \k) + \phi^{\rm DI}(\x).\label{eq:app_ddtec_f_a}
\end{align}
This functional form only specifies that the DI term is not dependent on $\k$.
The differential phase, to which the RIME is sensitive to, is found by spatially referencing the phase,
\begin{align}
    \Delta_0 \phi(\x, \k) =& \phi(\x, \k) - \phi(\x_0, \k)\\
    =&\Delta_0\phi^{\rm DD}(\x, \k) + \Delta_0 \phi^{\rm DI}(\x).\label{eq:app_ddtec_f_b}
\end{align}
Directionally referencing the differential phase to direction $\k_0$ we have,
\begin{align}
    \Delta_0^2 \phi(\x, \k) =& \Delta_0 \phi(\x, \k) - \Delta_0 \phi(\x, \k_0)\\
    =& \Delta_0^2\phi^{\rm DD}(\x, \k).
\end{align}
We see that the DI phase has disappeared, and we are left with doubly differential phase for the DD term.
Suppose then that there was a remnant DI component in $\phi^{\rm DD}(\x, \k)$.
Then by induction we have that $\Delta_0^2\phi^{\rm DD}(\x, \k)$ must be free of DI terms, and $\Delta_0^2 \phi(\x, \k)$ must be free of all DI components.
It follows that directionally referencing phase guarantees to remove all DI components.

Furthermore, suppose you were to directionally reference a doubly differential phase.
Suppose the doubly differential phase is referenced to direction $\k_0'$, with ${\Delta'}_0^2\phi(\x,\k) \triangleq \Delta_0 \phi(\x, \k) - \Delta_0 \phi(\x, \k_0')$.
Then we see,
\begin{align}
    {\Delta'}_0^2\phi(\x,\k) - {\Delta'}_0^2\phi(\x,\k_0) =& \Delta_0 \phi(\x, \k) - \Delta_0 \phi(\x, \k_0') - (\Delta_0 \phi(\x, \k_0) - \Delta_0 \phi(\x, \k_0'))\\
    =& \Delta_0 \phi(\x, \k)  - \Delta_0 \phi(\x, \k_0)\\
    =& \Delta_0^2 \phi(\x, \k).
\end{align}
That is, directionally referencing a doubly differential phase, produces a new doubly differential phase, referenced to the new direction, $\k_0$.
This trick thus allows one to set a well-defined reference direction to phases that have undergone DI calibration before DD calibration.

\section{Recursive Bayesian estimation}
\label{app:rbe}

\begin{wrapfigure}{r}{0.5\textwidth}
\centering
\def\svgwidth{0.5\columnwidth}
%% Accompanies image file 'hmm.pdf' (pdf, eps, ps)
\begingroup%
  \makeatletter%
  \providecommand\color[2][]{%
    \errmessage{(Inkscape) Color is used for the text in Inkscape, but the package 'color.sty' is not loaded}%
    \renewcommand\color[2][]{}%
  }%
  \providecommand\transparent[1]{%
    \errmessage{(Inkscape) Transparency is used (non-zero) for the text in Inkscape, but the package 'transparent.sty' is not loaded}%
    \renewcommand\transparent[1]{}%
  }%
  \providecommand\rotatebox[2]{#2}%
  \newcommand*\fsize{\dimexpr\f@size pt\relax}%
  \newcommand*\lineheight[1]{\fontsize{\fsize}{#1\fsize}\selectfont}%
  \ifx\svgwidth\undefined%
    \setlength{\unitlength}{267bp}%
    \ifx\svgscale\undefined%
      \relax%
    \else%
      \setlength{\unitlength}{\unitlength * \real{\svgscale}}%
    \fi%
  \else%
    \setlength{\unitlength}{\svgwidth}%
  \fi%
  \global\let\svgwidth\undefined%
  \global\let\svgscale\undefined%
  \makeatother%
  \begin{picture}(1,0.56460674)%
    \lineheight{1}%
    \setlength\tabcolsep{0pt}%
    \put(0,0){\includegraphics[width=\unitlength,page=1]{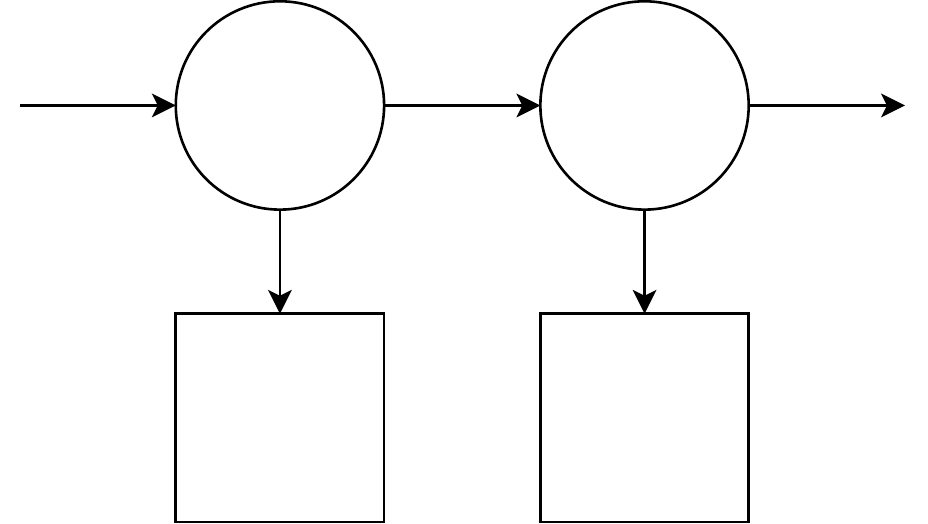}}%
    \put(0.64486994,0.44616261){\color[rgb]{0,0,0}\makebox(0,0)[lt]{\lineheight{84.55000305}\smash{\begin{tabular}[t]{l}\textbf{\textit{$\x_{i+1}$}}\end{tabular}}}}%
    \put(0.24203311,0.44337202){\color[rgb]{0,0,0}\makebox(0,0)[lt]{\lineheight{84.55000305}\smash{\begin{tabular}[t]{l}\textbf{\textit{$\x_{i}$}}\end{tabular}}}}%
    \put(0.23892911,0.10298663){\color[rgb]{0,0,0}\makebox(0,0)[lt]{\lineheight{84.55000305}\smash{\begin{tabular}[t]{l}\textbf{\textit{$\y_{i}$}}\end{tabular}}}}%
    \put(0.65385604,0.10292752){\color[rgb]{0,0,0}\makebox(0,0)[lt]{\lineheight{84.55000305}\smash{\begin{tabular}[t]{l}\textbf{\textit{$\y_{i+1}$}}\end{tabular}}}}%
  \end{picture}%
\endgroup%

\caption{Casual graph depicting a hidden Markov model.}
\label{fig:hmm}
\end{wrapfigure}
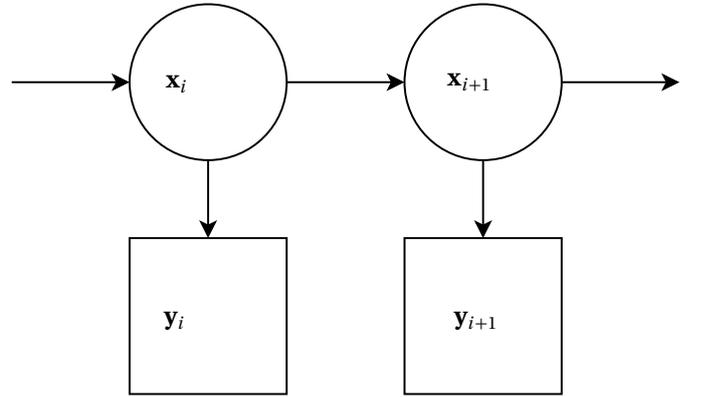

Recursive Bayesian estimation is a method of performing inference on a hidden Markov model \citep[HMM;][]{rabiner1986}.
Let $\y$ be an observable, and $\x$ be a hidden variable.
The HMM assumptions on $\x$ and $y$ are, firstly, that the hidden random variable is only conditionally dependent on its previous state, and secondly, the observable is conditionally independent of all other random variables except the the current hidden state.
These assumptions are depicted in \textbf{Figure}~\ref{fig:hmm}, where $i$ is the sequence index.
This paradigm is often given a notion of causality, or time, however this is not necessary in any way.
The sequence index is an abstract notion, that simply explains how the set of hidden variables are traversed.
For example, suppose the observations were frames of a movie, and the hidden variable was the plot contained in each movie frame.
The HMM assumption are that the movie plot is linear, and the picture encodes what's going on in the movie at a given point in time.
With recursive Bayesian estimation you could watch the movie with frames randomly ordered and still get the complete plot.

There are two distinct types of information propagation in a hidden Markov model.
Information can flow in the direction of the arrows, or against them.
This leads to the notion of the forward and backwards equations which describe how to propagate belief in hidden variables forward, and for revising you belief in previously visited hidden variables respectively.

The joint distribution of the hidden random variables and observables in a chain of length $T$ can be written out as a product of conditional distributions and a marginal using the product rule of probability distributions \citep{kolmogorov1960foundations}.
Because the HMM assumptions the joint distribution is,
\begin{align}
p(\x_{0:T}, \y_{0:T}) =& p(\x_0) \prod_{i=1}^T p(\x_{i} \mid \x_{i-1}) p(\y_{i} \mid \x_{i}).
\end{align}

Let's consider first how to propagate information forward.
This is done in two steps, typically called the predict and update steps.
For the predict step, we consider how to propagate belief in the absence of new observables.
For this, we apply the Chapman-Kolmogorov identity\footnote{If $b$ is conditionally independent of $a$ then $p(a\mid c) = \int p(a\mid b)p(b | c)\,\mathrm{d}b = \mathbb{E}_{b | c}[p(a\mid b)]$ is the Chapman-Kolmogorov identity.} for Markovian processes,
\begin{align}
 p(\x_{i+1} \mid \y_{0:i}) =& \mathbb{E}_{\x_{i} \mid \y_{0:i}}\left[p(\x_{i+1} \mid \x_{i}) \right]\label{eq:predict}
\end{align}
which gives us the probability distribution of the hidden variables at time $i+1$ in terms of the so-called state transition distribution $p(\x_{i +1} \mid \x_{i})$ and posterior distribution $p(\x_{i} \mid \y_{0:i})$ at index $i$.
The current prior belief can be understood as the expectation of the state transition distribution over the measure of the current posterior belief.

The update step is simply an application of Bayes theorem with the prior defined by the predict step,
\begin{align}
p(\x_{i} \mid \y_{0:i}) =& \frac{p(\y_{i} \mid \x_{i}) \mathbb{E}_{\x_{i-1} \mid \y_{0:i-1}}\left[p(\x_{i} \mid \x_{i-1}) \right]}{p(\y_{i} \mid \y_{0:i-1})},\label{eq:update}
\end{align}
where the denominator is the Bayesian evidence of the newly arrived data given all previous data, and it is independent of the hidden variables.
Equation~\ref{eq:update} gives a recurrence relation for propagating our belief forward, therefore this is called the forward equation.

Suppose now that you at at index $T$, and wish to use all acquired information to revise your belief in the previously visited hidden variables at indices $i<T$.
The trick is to realise that $p(\x_i \mid \x_{i+1} \mid \y_{0:T}) = p(\x_i \mid \x_{i+1} \mid \y_{0:i})$ due to the Markov properties.
In this case, again using the product rule, we find that the joint conditional distribution of a pair of sequential hidden states given the whole sequence of data is,
\begin{align}
    p(\x_i, \x_{i+1} \mid \y_{0:T}) = \frac{p(\x_{i+1}\mid \x_i) p(\x_{i} \mid \y_{0:i}) p(\x_{i+1} \mid \y_{0:T})}{p(\x_{i+1} \mid \y_{0:i})}.
\end{align}
Marginalising the second hidden parameter we arrive at the recurrence relation,
\begin{align}
    p(\x_i \mid \y_{0:T}) = p(\x_{i} \mid \y_{0:i}) \int \frac{p(\x_{i+1}\mid \x_i)  p(\x_{i+1} \mid \y_{0:T})}{p(\x_{i+1} \mid \y_{0:i})}\, \mathrm{d}\x_{i+1}.\label{eq:backwards}
\end{align}
This can be solved by starting at $T$ and solving this equation iteratively backwards, therefore Equation~\ref{eq:backwards} is called the backward equation.
Most importantly, note that the backward equation does not require conditioning on data as was done in the update step.

When the transition and likelihood are assumed to be Normal, e.g. in a linear dynamical system, the forward and backward equations are equivalent to the well-known Kalman filter equations and Rauch smoother equations \citep{rauch1963} respectively.

\section{Jones scalar variational expectation}
\label{app:var_exp}

Let $\g \in \mathbb{C}^{N_{\rm freq}}$ be an observed complex Jones scalar vector, with amplitudes $g \in \mathbb{R}^{N_{\rm freq}}$ and phases $\phi \in \mathbb{R}^{N_{\rm freq}}$.
We assume that the Jones scalars have complex Gaussian noise, described by the observational covariance matrix $\bs{\Sigma}$.
Thus, we have that the observational likelihood of the Jones scalars (\textit{cf} Eq.~\ref{eq:vi_likelihood}) is,
\begin{align}
    p(\g \mid \phi, g, \bs{\Sigma}) =& \mathcal{N}_\mathbb{C}[\g \mid (g \cos{ \phi}, g \sin \phi)^T, \bs{\Sigma}],
\end{align}
where $\mathcal{N}_\mathbb{C}$ is the complex Gaussian distribution, which is defined as the Gaussian distribution of the stacked real and imaginary components.
Let us define the residuals $\delta R = \mathrm{Re}[\g] - g \cos \phi$ and $\delta I = \mathrm{Im}[\g] - g \sin \phi$, and the stacked residuals $\delta \g = (\delta R, \delta I)^T$.
Then the log-likelihood becomes,
\begin{align}
    \log p(\g \mid \phi, g, \bs{\Sigma}) =& -N_{\rm freq} \log 2 \pi - \log|\bs{\Sigma}| - \frac{1}{2} \delta \g^T \bs{\Sigma}^{-1} \delta \g\\
    =& -N_{\rm freq} \log 2 \pi - \log|\bs{\Sigma}| - \frac{1}{2} \mathrm{Tr}[\bs{\Sigma}^{-1} \delta \g \delta \g^T]\\
    =& -N_{\rm freq} \log 2 \pi - \log|\bs{\Sigma}| - \frac{1}{2} \mathrm{vec}[\bs{\Sigma}^{-1}]^{T} \mathrm{vec}[\delta \g \delta \g^T]
\end{align}
where we used the fact that the trace of a scalar is a scalar, and that $\mathrm{Tr}[A^T B] = \mathrm{Tr}[B A^T]=\mathrm{vec} [A]^T \mathrm{vec} [B]$.

Now let us suppose that the phases are linearly modelled according to,
\begin{align}
    \phi(\nu) = \sum_i^{M} f_i(\nu) a_i,
\end{align}
where $f_i(\nu)$ is the $i$-th basis function of a set of $M$ linearly independent functions depending on frequency $\nu$, and $a_i$ is the $i$-th parameter in the linear model.
Let us further suppose that each parameter $a_i$ is a Gaussian random variable,
\begin{align}
    a_i \sim \mathcal{N}[m_i, l_i^2],
\end{align}
and that all parameters are independent, i.e. $p(a_i, a_j) = p(a_i)p(a_j)$.
Because all parameters are independent and the basis functions are linearly independent the phase is the unique Gaussian random variable with distribution, 
\begin{align}
    p(\phi(\nu)) = \mathcal{N}\left[\sum_i^M f_i(\nu) m_i, \sum_i^M f^2_i(\nu) l_i^2\right].
\end{align}
The variational expectation of the Jones scalars with respect to the phase is thus defined as, 
\begin{align}
    \mathbb{E}_{p(\phi)}\left[\log p(\g \mid \phi, g, \bs{\Sigma}) \right]
    =&
    -N_{\rm freq} \log 2 \pi - \log|\bs{\Sigma}| - \frac{1}{2} \mathrm{vec}[\bs{\Sigma}]^T \mathrm{vec}\left[\mathbb{E}_{p(\phi)}[\delta \g \delta \g^T]\right],\label{eq:var_exp_anal_a}
\end{align}
where we used that expectation and the vectorisation operator commute.
Now the only thing that needs to be evaluated is the expectation, $\mathbb{E}_{p(\phi)}\left[ \delta \g \delta \g^T\right]$.
In order to do this, we realise that there are three explicit integrations that must be done corresponding to unique sub-blocks of $\delta \g \delta \g^T$ consisting of the real-real, real-imaginary, and imaginary-imaginary outer products of residual vectors,
\begin{align}
    \mathbb{E}_{p(\phi)}\left[\delta R \delta R^T\right] =&  \int \delta R \delta R^T\, \mathrm{d}p(\phi)\\
    \triangleq&   I_{rr}\\
     \mathbb{E}_{p(\phi)}\left[ \delta R \delta I^T\right] =&  \int \delta R \delta I^T\, \mathrm{d}p(\phi)  \\
    \triangleq&  I_{ri}\\
     \mathbb{E}_{p(\phi)}\left[\delta I \delta I^T\right] =& \int \delta I\delta I^T\, \mathrm{d}p(\phi)\\
    \triangleq&   I_{ii}.
\end{align}
All three of the above integrals involve Gaussian expectations of trigonometric functions.
Specifically, consider the $(i,j)$-th element of the integral $I_{rr}$,
\begin{align}
    (I_{rr})_{(i,j)}  =& \int (\mathrm{Re}[\g]_i - g_i \cos \phi(\nu_i)) (\mathrm{Re}[\g]_j - g_j \cos \phi(\nu_j))\, \mathrm{d}p(\phi(\nu_i), \phi(\nu_j)).
\end{align}

To evaluate this integral use Euler's formulae, e.g. for cosine, $2 \cos \phi = e^{\iota \phi} + e^{-\iota \phi}$.
This allows us to reduce all combinations of trigonometric functions to complex exponentials.
We then use the characteristic function of the Gaussian distribution which gives the relation, $\mathbb{E}_{X \sim \mathcal{N}[m, l^2]}[e^{\iota X}] = e^{\iota m - \frac{l^2}{2}}$.
Applying the above, after significant elementary algebraic work we evaluate all integral matrix elements,
\begin{align}  
    (I_{rr})_{(i,j)} =& 
    \left(\mathrm{Re}[\g]_i - e^{-\frac{1}{2}\sum_k^M l_k^2 f^2_k(\nu_i)} g_i \cos \left(\sum_k^M m_k f_k(\nu_i)\right) \right) 
    \left(\mathrm{Re}[\g]_j - e^{-\frac{1}{2}\sum_k^M l_k^2  f^2_k(\nu_j)} g_j \cos \left(\sum_k^M m_k f_k(\nu_j)\right)\right)\\
    (I_{ri})_{(i,j)} =& \left(\mathrm{Re}[\g]_i - e^{-\frac{1}{2}\sum_k^M l_k^2 f^2_k(\nu_i)} g_i \cos \left(\sum_k^M m_k f_k(\nu_i)\right) \right) 
    \left(\mathrm{Im}[\g]_j - e^{-\frac{1}{2}\sum_k^M l_k^2  f^2_k(\nu_j)} g_j \sin \left(\sum_k^M m_k f_k(\nu_j)\right)\right)\\
    (I_{ii})_{(i,j)} =& \left(\mathrm{Im}[\g]_i - e^{-\frac{1}{2}\sum_k^M l_k^2 f^2_k(\nu_i)} g_i \sin \left(\sum_k^M m_k f_k(\nu_i)\right) \right) 
    \left(\mathrm{Im}[\g]_j - e^{-\frac{1}{2}\sum_k^M l_k^2  f^2_k(\nu_j)} g_j \sin \left(\sum_k^M m_k f_k(\nu_j)\right)\right)
\end{align}
This completes the evaluation of Eq.~\ref{eq:var_exp_anal_a} for arbitrary linear phase model.
For example, in a phase model consisting only of DDTEC we have $M=1$, $f_1(\nu) = \frac{\kappa}{\nu}$ and $a_1 = \Delta_0^2 \tau$.
It is simple to calculate the variational expectation of other linear phase models, e.g. clock-like terms, constant-in-frequency terms, higher order ionospheric terms.

\end{document}